\documentclass[submission, Phys]{SciPost}

\usepackage[utf8]{inputenc} 
\usepackage[T1]{fontenc} 	
\usepackage[english]{babel} 

\usepackage[bitstream-charter]{mathdesign}
\usepackage{geometry} 		
\usepackage{amsmath} 		
\usepackage{mathtools} 		
\usepackage{float} 			
\usepackage{graphicx} 		
\usepackage{tabularx} 		
\usepackage{booktabs} 		
\usepackage{color, xcolor} 	
\usepackage{pdfpages} 		
\usepackage{extarrows} 		
\usepackage{multirow} 		
\usepackage{multicol} 		
\usepackage{enumitem} 		
\usepackage{xspace} 		
\usepackage{stackrel} 		
\usepackage{tikz} 			
\usepackage{braket} 		
\usepackage{bm} 			
\usepackage{tensor} 		
\usepackage{slashed} 		
\usepackage{siunitx} 		
\usepackage{lastpage} 		
\usepackage{cite} 			
\usepackage[normalem]{ulem} 
\usepackage{fontawesome} 	
\usepackage{tocloft} 		
\usepackage{titlesec} 		
\usepackage{doi} 			
\usepackage{hyperref} 		
\usepackage[most]{tcolorbox} 					
\usepackage[nameinlink, capitalize]{cleveref} 	
\usepackage[nottoc, notlot, notlof]{tocbibind} 	
\usepackage[ruled, vlined]{algorithm2e} 		
\usepackage{makecell}
\usepackage[makeroom]{cancel}
\usepackage{feynmf}

\makeatletter
\def\BState{\State\hskip-\ALG@thistlm}
\makeatother

\makeatletter
\@ifundefined{pdfoutput}{}{\DeclareGraphicsRule{*}{mps}{*}{}}
\makeatother

\makeatletter
\DeclareRobustCommand*{\bfseries}{%
   \not@math@alphabet\bfseries\mathbf
   \fontseries\bfdefault\selectfont
   \boldmath
}
\makeatother

\hypersetup{
	pdftitle={MadNIS},
	pdfauthor={Heimel et al.},
	colorlinks=true, 			
	linkcolor={red!50!black}, 	
	citecolor={blue!50!black}, 	
	urlcolor={blue!80!black} 	
} 

\DeclareSymbolFont{usualmathcal}{OMS}{cmsy}{m}{n}
\DeclareSymbolFontAlphabet{\mathcal}{usualmathcal}

\SetArgSty{textnormal}
\SetKwComment{Comment}{{\small\#}~}{}
\SetCommentSty{mycommfont}

\setitemize{itemsep=0pt, parsep=0pt} 				
\setenumerate{itemsep=0pt, parsep=0pt} 				
\setitemize{itemsep=2pt,topsep=2pt,parsep=0pt,partopsep=0pt,leftmargin=*}
\setenumerate{itemsep=0pt,topsep=2pt,parsep=0pt,partopsep=0pt,labelindent=3pt,leftmargin=*}
\usepackage{amsmath}
 
\usepackage{amsthm} 		
\theoremstyle{definition}
 
\definecolor{red_cb}{HTML}{e41a1c}
\definecolor{blue_cb}{HTML}{377eb8}
\definecolor{green_cb}{HTML}{4daf4a}
\definecolor{purple_cb}{HTML}{984ea3}
\definecolor{orange_cb}{HTML}{ff7f00}

\definecolor{EmeraldGreen}{HTML}{1ea78d}
\definecolor{EnglishRed}{HTML}{b02427}
\hypersetup{colorlinks=true,urlcolor=EmeraldGreen,citecolor=EmeraldGreen,linkcolor=EnglishRed}

\newcommand{\ie}{\text{i.e.}\;}

\newcommand{\ai}{\alpha_i}
\newcommand{\aix}{\alpha_{i \xi}}

\newcommand{\Git}{G_{i \theta}}
\newcommand{\git}{g_{i \theta}}
\newcommand{\gbar}{\overline{G}}

\newcommand{\pl}{p_0}

\newcommand{\mwith}{\text{with}}
\newcommand{\mand}{\text{and}}
\newcommand{\mfor}{\text{for}}

\newcommand{\qqqquad}{\qquad\qquad}

\def\d{\mathrm{d}}

\newcommand\one{\leavevmode\hbox{\small1\normalsize\kern-.33em1}}
\newcommand{\del}{\partial} 				
\newcommand{\imag}{\mathrm{i}} 				

\newcommand{\loss}{\mathcal{L}} 	

\newcommand{\vegas}{\textsc{Vegas}\xspace}

\newcommand{\mg}{\textsc{MG5aMC}\xspace}
\newcommand{\pythia}{\textsc{Pythia8}\xspace}

\newcommand{\tensorflow}{\textsc{TensorFlow}\xspace}

\newcommand{\pytorch}{\textsc{PyTorch}\xspace}
\newcommand{\sherpa}{\textsc{Sherpa}\xspace}

\newcommand{\madnis}{\textsc{MadNIS}\xspace}

\newcommand{\whizard}{\textsc{Whizard}\xspace}

\newcommand{\rambodiet}{\textsc{RamboOnDiet}\xspace}
\newcommand{\pdfflow}{\textsc{PDFFlow}\xspace}
\newcommand{\madflow}{\textsc{MadFlow}\xspace}
\newcommand{\madjax}{\textsc{MadJax}\xspace}

\newcommand{\arXiv}[2][]{%
	\ifthenelse{\equal{#1}{}}%
	{\href{http://arxiv.org/abs/#2}{arXiv:#2}}%
	{\href{http://arxiv.org/abs/#2}{arXiv:#2~[#1]}}}

\def\slashchar#1{\setbox0=\hbox{$#1$}           
   \dimen0=\wd0                                 
   \setbox1=\hbox{/} \dimen1=\wd1               
   \ifdim\dimen0>\dimen1                        
      \rlap{\hbox to \dimen0{\hfil/\hfil}}      
      #1                                        
   \else                                        
      \rlap{\hbox to \dimen1{\hfil$#1$\hfil}}   
      /                                         
   \fi}

\newcommand{\tikznode}[2]{%
\ifmmode%
\tikz[remember picture,baseline=(#1.base),inner sep=0pt] \node (#1) {$#2$};%
\else
\tikz[remember picture,baseline=(#1.base),inner sep=0pt] \node (#1) {#2};%
\fi}

\def\mathswitchr#1{\relax\ifmmode{\mathrm{#1}}\else$\mathrm{#1}$\xspace\fi}
\def\mathswitch#1{\relax\ifmmode#1\else$#1$\xspace\fi}

\newcommand{\PW}{\mathswitchr W}
\newcommand{\PWp}{\mathswitchr {W^+}}
\newcommand{\PWm}{\mathswitchr {W^-}}

\newcommand{\Pg}{\mathswitchr g}

\newcommand{\Pe}{\mathswitchr e}

\newcommand{\Pd}{\mathswitchr d}
\newcommand{\Pu}{\mathswitchr u}

\newcommand{\Pt}{\mathswitchr t}

\newcommand{\Ptbar}{\mathswitchr{\bar t}}
\newcommand{\Pubar}{\mathswitchr{\bar u}}
\newcommand{\Pdbar}{\mathswitchr{\bar d}}

\newcommand{\jets}{\mathrm{jets}}

\graphicspath{{./figs/}}

\begin{document}

\begin{center}{\Large \textbf{
Differentiable \madnis-Lite}}
\end{center}

\begin{center}
Theo Heimel\textsuperscript{1},
Olivier Mattelaer\textsuperscript{2},
Tilman Plehn\textsuperscript{1,3},
and Ramon Winterhalder\textsuperscript{2}
\end{center}

\begin{center}
{\bf 1} Institut f\"ur Theoretische Physik, Universit\"at Heidelberg, Germany
\\
{\bf 2} CP3, Universit\'e catholique de Louvain, Louvain-la-Neuve, Belgium \\
{\bf 3} Interdisciplinary Center for Scientific Computing (IWR), Universit\"at Heidelberg, Germany
\end{center}

\begin{center}
\today
\end{center}


\section*{Abstract}
{\bf Differentiable programming opens exciting new avenues in particle physics, also affecting future event generators. These new techniques boost the performance of current and planned MadGraph implementations. Combining phase-space mappings with a set of very small learnable flow elements, \madnis-Lite, can improve the sampling efficiency while being physically interpretable. This defines a third sampling strategy, complementing \vegas and the full \madnis.}

\vspace{10pt}
\noindent\rule{\textwidth}{1pt}
\tableofcontents\thispagestyle{fancy}
\noindent\rule{\textwidth}{1pt}
\vspace{10pt}

\clearpage
\section{Introduction}
\label{sec:intro}

The incredibly successful precision-LHC program is based on comparing vast numbers of scattering events with first-principle predictions provided by multi-purpose event generators, like \pythia~\cite{Sjostrand:2014zea},
\mg~\cite{Alwall:2014hca}, and \sherpa~\cite{Sherpa:2019gpd}. 
For a given Lagrangian, they use perturbative quantum
field theory to provide the backbone of a complex simulation and
inference chain. In view of the upcoming LHC runs, we will have to rely on both modern machine learning (ML)~\cite{Butter:2022rso,Plehn:2022ftl} and improved hardware accelerated codes~\cite{Bothmann:2022thx,Valassi:2023yud,Bothmann:2023gew,Hageboeck:2023blb,Wettersten:2023ekm} to significantly improve the speed and the precision of these simulations. The \madnis~\cite{Heimel:2022wyj,Heimel:2023ngj} framework is a promising candidate for such ML-based advancements.
Following the modular structure of event generators, 
modern neural networks will transform phase-space sampling~\cite{Bendavid:2017zhk,Klimek:2018mza,Chen:2020nfb,Gao:2020vdv,Bothmann:2020ywa,Gao:2020zvv,Danziger:2021eeg,Heimel:2022wyj,Janssen:2023ahv,Bothmann:2023siu,Heimel:2023ngj, Deutschmann:2024lml}, scattering amplitude
evaluations~\cite{Bishara:2019iwh,Badger:2020uow,Aylett-Bullock:2021hmo,Maitre:2021uaa,Winterhalder:2021ngy,Badger:2022hwf,Maitre:2023dqz}, event generation~\cite{Otten:2019hhl,Hashemi:2019fkn,DiSipio:2019imz,Butter:2019cae,Alanazi:2020klf,Butter:2021csz,Butter:2023fov},
parton shower generation~\cite{deOliveira:2017pjk,Andreassen:2018apy,Bothmann:2018trh,Dohi:2020eda,Buhmann:2023pmh,Leigh:2023toe,Mikuni:2023dvk,Buhmann:2023zgc}, as well as the critical and currently far too slow detector
simulations~\cite{Paganini:2017hrr,deOliveira:2017rwa,Paganini:2017dwg,Erdmann:2018kuh,Erdmann:2018jxd,Belayneh:2019vyx,Buhmann:2020pmy,Buhmann:2021lxj,Krause:2021ilc,
  ATLAS:2021pzo,Krause:2021wez,Buhmann:2021caf,Chen:2021gdz,
  Mikuni:2022xry,ATLAS:2022jhk,Krause:2022jna,Cresswell:2022tof,Diefenbacher:2023vsw,
Hashemi:2023ruu,Xu:2023xdc,Diefenbacher:2023prl,Buhmann:2023bwk,Buckley:2023daw,Diefenbacher:2023flw,Ernst:2023qvn,Favaro:2024rle,Buss:2024orz,Quetant:2024ftg
}.

The workhorses for this transformation are generative networks, trained on and amplifying simulated training data~\cite{Butter:2020qhk,Bieringer:2022cbs}. Given the LHC requirements, they have to 
be controlled and precise in encoding kinematic patterns over an, essentially, interpretable phase space~\cite{Butter:2021csz,
  Winterhalder:2021ave,Nachman:2023clf,Leigh:2023zle,Das:2023ktd}. In addition to 
event generation, these 
generative networks can be used for event
subtraction~\cite{Butter:2019eyo}, event
unweighting~\cite{Stienen:2020gns,Backes:2020vka}, or
super-resolution enhancement~\cite{DiBello:2020bas,Baldi:2020hjm}.
Their conditional versions enable new analysis methods, like
probabilistic
unfolding~\cite{Datta:2018mwd,Bellagente:2019uyp,Andreassen:2019cjw,Bellagente:2020piv,Backes:2022vmn,Leigh:2022lpn,Raine:2023fko,Shmakov:2023kjj,Ackerschott:2023nax,Diefenbacher:2023wec},
inference~\cite{Bieringer:2020tnw,Butter:2022vkj,Heimel:2023mvw,Du:2024gbp}, or
anomaly
detection~\cite{Nachman:2020lpy,Hallin:2021wme,Raine:2022hht,Hallin:2022eoq,Golling:2022nkl,Sengupta:2023xqy}.

An even more advanced strategy to improve LHC simulations is differentiable programming. Here, the entire simulation chain is envisioned 
to benefit from the availability of derivatives with respect to, for instance, model and tuning parameters in modern computer languages. A proof of principle has 
been delivered for 
differentiable matrix elements~\cite{Heinrich:2022xfa},
but the same methods are used for differentiable detector design~\cite{MODE:2022znx}, derivatives of branching processes~\cite{Kagan:2023gxz},
shower-simulations with path-wise derivatives~\cite{Aehle:2024ezu}, all the way to 
a differentiable parton-shower event generator for $\Pe^+\Pe^-$ collisions~\cite{Nachman:2022jbj}.
The crucial question is where differentiable programming can be used to improve existing
ML-enhanced classic event generators.

We will test how a differentiable version of the 
\madnis event sampler~\cite{Heimel:2022wyj,Heimel:2023ngj} compares to established 
improvements which will be part of an upcoming \textsc{MadGraph} release for the HL-LHC.
We expect our findings to similarly apply to neural importance sampling (NIS) developments in \sherpa~\cite{Gao:2020vdv,Bothmann:2020ywa}.
In Sec.~\ref{sec:madnis} we briefly review the \madnis reference structures, into which 
we first implement a differentiable integrand, including phase space, matrix element, and parton densities, in Sec.~\ref{sec:integrand}. 
In a second step, we construct a differentiable phase space generator to improve the
phase space mapping through a set of small learnable elements, \madnis-Lite, in Sec.~\ref{sec:phase_space}. 
Here we find that differentiable
programming can be useful for ML-event generators, but the performance
gain has to be benchmarked and evaluated carefully. For \madnis-Lite, 
a differentiable phase space generator appears very promising.

\section{\madnis basics}
\label{sec:madnis}

We briefly review multi-channel Monte Carlo and the \madnis basics~\cite{Heimel:2022wyj,Heimel:2023ngj}, necessary to understand the subsequent sections. We consider the integral of a function $f
\sim \vert\mathcal{M}\vert^2$ over phase space
\begin{align}
    I[f] = \int \d x\,f(x)
    \qqqquad
    x \in \mathbb{R}^D \; .
    \label{eq:psinteg}
\end{align}
It can be decomposed by introducing local channel weights
$\ai(x)$~\cite{Maltoni:2002qb,Mattelaer:2021xdr}
\begin{align}
  f(x)
  = \sum^{n_c}_{i=1} \ai(x) \; f(x)
  \qquad \mwith \quad \sum^{n_c}_{i=1} \ai(x) =1
  \quad \mand \quad \ai(x)\ge 0\;,
\label{eq:norm_alpha}
\end{align}
using the \mg decomposition. Similar decompositions~\cite{Kleiss:1994qy, Weinzierl:2000wd} are used in 
\sherpa~\cite{Sherpa:2019gpd} and \whizard~\cite{Kilian:2007gr}. The phase-space integral now
reads
\begin{align}
  I[f]
  = \sum^{n_c}_{i=1}\int \d x\;\ai(x)\,f(x) \; .
  \label{eq:multi-channel-mg1}
\end{align}
Next, we introduce a set of channel-dependent phase-space mappings
\begin{align}
x \in \mathbb{R}^D \quad 
\xleftrightarrow[\quad \leftarrow \gbar_i(z)\quad]{G_i(x)\rightarrow} 
\quad z\in[0,1]^D  \; ,
\label{eq:ps_mapping}
\end{align}
which parametrize properly normalized densities
\begin{align}
   g_i(x)=\left\vert\frac{\partial G_i(x)}{\partial x}\right\vert
   \qquad \mwith \quad 
   \int \d x\,g_i(x)=1  \; .
   \label{eq:channel_densities}
\end{align}
The phase-space integral now covers the $D$-dimensional unit hypercube and is sampled as
\begin{align}
\begin{split}
  I[f]
  &= \sum^{n_c}_{i=1}\int \left. \frac{\d z}{g_i(x)} \; \alpha_i(x) f(x) \right\vert_{x=\gbar_i(z)} \\
  &= \sum^{n_c}_{i=1}\int \d x \; g_i(x) \; \frac{\alpha_i(x) f(x)}{g_i(x)}
  \equiv \sum^{n_c}_{i=1} \left\langle \frac{\alpha_i(x) f(x)}{g_i(x)} \right\rangle_{x\sim g_i(x)}  \; .
\end{split}
  \label{eq:multi-channel-mg2}
\end{align}
This is the basis of multi-channel importance sampling~\cite{Weinzierl:2000wd}. Starting from this equation, \madnis encodes the multi-channel weight $\ai(x)$ and the channel mappings $G_i(x)$ in
neural networks. 

\subsubsection*{Neural channel weights}

First, \madnis employs a channel-weight network to
encode the local multi-channel weights 
\begin{align}
   \ai(x)\equiv\aix(x) \; ,
  \label{eq:trained_weights}
\end{align}
where $\xi$ denotes the network parameters. 
As the channel weights vary strongly over phase space, it helps the performance to learn them as a correction to a physically motivated prior assumption like~\cite{Maltoni:2002qb,Mattelaer:2021xdr} 
\begin{align}
\begin{split}
    \ai^{\text{MG},1}(x)&= \frac{|\mathcal{M}_i(x)|^2}{\sum_j |\mathcal{M}_j(x)|^2}\;, \quad \text{or} \\
    \ai^{\text{MG},2}(x) &= \frac{P_i(x)}{\sum_j P_j(x)} 
  \quad \text{with} \quad
  P_i(x) =  \prod_{k \in \text{prop}} \frac{1}{|p_k(x)^2-m_k^2 -\imag m_k\Gamma_k|^2} \; .
\end{split}
\label{eq:mg_prior}
\end{align}
Relative to either of them, we then only learn a correction factor~\cite{Heimel:2023ngj}.

\subsubsection*{Neural importance sampling}

Second, \madnis combines analytic channel mappings with a normalizing flow~\cite{inn,Kobyzev_2020},
\begin{align}
x \in \mathbb{R}^D 
\xleftrightarrow[\hspace{1.1cm}]{\text{analytic}} 
y\in[0,1]^D
\xleftrightarrow[\hspace{1.1cm}]{\text{flow}}
z\in[0,1]^D \; ,
\label{eq:def_inn}
\end{align}
to complement \vegas~\cite{Lepage:1977sw,Lepage:1980dq,Lepage:2020tgj}. The flow allows 
\madnis to improve
the physics-inspired phase-space mappings by training a network to map
\begin{align}
  z = \Git(x) \qquad \text{or} \qquad x = \gbar_{i \theta}(z) \; ,
  \label{eq:trained_mappings}
\end{align} 
where $\theta$ denotes another set of network parameters.
As for the channel weights, we use physics knowledge to simplify the ML task and a \vegas pre-training~\cite{Heimel:2023ngj}.
This makes use of the key strength of \vegas,
which is extremely efficient for factorizing integrands and converges much faster than neural importance sampling.

\subsubsection*{Multi-channel variance loss}

With the channel weights and importance sampling encoded in neural networks,
\begin{align}
\ai(x) \equiv \aix(x)
\qquad \mand \qquad 
g_i(x)\equiv \git(x) \; ,
\end{align}
the integral in Eq.\eqref{eq:multi-channel-mg2} becomes
\begin{align}
  I[f]
  = \sum^{n_c}_{i=1}\left\langle\frac{\aix(x) f(x)}{\git(x)}\right\rangle_{x\sim \git(x)} \; .
\end{align}
Crucially, we then minimize the variance as the loss~\cite{Heimel:2022wyj,Heimel:2023ngj}
\begin{align}
\begin{split}
\loss_\text{variance}
&=\sum^{n_c}_{i=1} \frac{N}{N_i}\sigma^2_i \\
&=\sum^{n_c}_{i=1}\frac{N}{N_i}\left(
    \left\langle \frac{\aix(x)^2 f(x)^2}{\git(x)\,q_i(x)} \right\rangle_{x\sim q_i(x)}
      - \left\langle \frac{\aix(x) f(x)}{q_i(x)} \right\rangle_{x\sim q_i(x)}^2 \right) \;.
\end{split}
\label{eq:var_loss}
\end{align}
In practice, $q_i(x)\simeq \git(x)$ allows us to compute the loss as
precisely as possible and stabilizes the combined
online~\cite{Butter:2022lkf} and buffered
training~\cite{Heimel:2022wyj}.
Inspired by stratified sampling, we encode the optimal choice for $N_i$~\cite{Press:1989vk,Weinzierl:2000wd} in the \madnis loss~\cite{Heimel:2023ngj}
\begin{align}
\begin{split}
    \loss_\madnis 
    &= \sum_{i=1}^{n_c} \left(\sum_{j=1}^{n_c}\sigma_j\right)\sigma_i= \left[\sum_{i=1}^{n_c} \sigma_i\right]^2 \\
    &= \left[\sum_{i=1}^{n_c} \left(
    \left\langle \frac{\aix(x)^2 f(x)^2}{\git(x)\,q_i(x)} \right\rangle_{x\sim q_i(x)}
      - \left\langle \frac{\aix(x) f(x)}{q_i(x)} \right\rangle_{x\sim q_i(x)}^2 \right)^{1/2} \right]^2 
    \;.
\end{split}
    \label{eq:madnis_loss}
\end{align}
%

\section{Differentiable integrand}
\label{sec:integrand}

To investigate how the choice of the loss function and the direction of the training affects the performance of \madnis, we consider a realistic LHC process, namely triple-W production,
\begin{align}
\Pu \Pdbar \to \PWp \PWp \PWm
\label{eq:www_proc_diff}  
\end{align}
At leading order, this process comes with 17 Feynman diagrams and 16 integration channels in \mg. It has been shown to benefit significantly from ML-based importance sampling methods~\cite{Heimel:2023ngj} and that a single fine-tuned integration channel is sufficient to achieve good performance. We have implemented a simple differentiable event generator for this process in \madnis, including
\begin{itemize}
    \item a differentiable squared matrix element using helicity amplitudes natively written in \pytorch that are generated from our custom \mg plugin similar to \madflow~\cite{Carrazza:2021gpx} and \madjax~\cite{Heinrich:2022xfa};

    \item a differentiable and invertible phase-space generator based on \rambodiet~\cite{Rambo,RamboDiet,Nachman:2023clf} natively written in \pytorch. A similar implementation has been used in \madjax~\cite{Heinrich:2022xfa};

    \item and differentiable parton densities using the standard LHAPDF interpolation~\cite{Whalley:2005nh} natively written in \pytorch which is similar to \pdfflow~\cite{Carrazza:2020qwu} relying on \tensorflow.   
\end{itemize}

\subsection{Forward and inverse training}

As a starting point we consider a generic $F$-divergence~\cite{Nielsen:2014}  between two normalized probability distributions $p_1(z)$ and $p_2(z)$,
\begin{align}
    D_F^z[p_1,p_2] = \int \d z\,p_2(z) \: F\!\left(\frac{p_1(z)}{p_2(z)}\right) \;.
\end{align}
For a \madnis-like training, we have a target function $f(x)$ and a normalizing flow that parametrizes a trainable invertible mapping with network parameters $\theta$
\begin{align}
x \quad 
\xleftrightarrow[\quad \leftarrow \gbar_\theta(z)\quad]{G_\theta(x)\rightarrow} 
\quad z
\label{eq:ps_mapping2}
\end{align}
and induces the density distribution
\begin{align}
    g_\theta(x) = \pl(G_\theta(x)) \left| \frac{\del G_\theta(x)}{\del x} \right| \; ,
\label{eq:forward_prob}
\end{align}
with latent space distribution $p_0(z)$.
For convenience, we can also define
\begin{align}
    \overline{g}_\theta(z) = \pl(z) \left| \frac{\del \gbar_\theta(z)}{\del z} \right|^{-1} 
    \quad \text{such that} \quad 
    \overline{g}_\theta(G_\theta(x)) = g_\theta(x) \;.
\end{align}
Note that $g_\theta(x)$ is a normalized probability distribution in $x$-space (data space), but this is not the case for $\overline{g}_\theta(z)$ in $z$-space (latent space). There are two ways to define a loss function to train the flow. First, we define the loss function in data space,
\begin{align}
\begin{split}
    \loss^\text{fw}_F = D_F^x[f,g_\theta]
    = \int \d x\,g_\theta(x) \: F\!\left(\frac{f(x)}{g_\theta(x)}\right)
    = \left\langle \frac{g_\theta(x)}{q(x)} \: F\!\left(\frac{f(x)}{g_\theta(x)} \right) \right\rangle_{x \sim q(x)} \; .
\end{split}
\end{align}
In the last step, we introduce an importance sampling distribution $q(x)$ to evaluate the integral numerically. This can be the same as $g_\theta(x)$ for online training, or different for buffered training~\cite{Heimel:2023ngj}. Optimizing this loss function requires evaluating the flow in the forward direction according to Eq.\eqref{eq:forward_prob}, so we refer to this training mode as forward training.

Alternatively, we can train in latent space using the remapped target distribution
\begin{align}
    \hat{f}(z) = f\!\left(\gbar_\theta(z)\right)\,\left\vert \frac{\del \gbar_\theta(z)}{\del z} \right\vert \; ,
\end{align}
which is a normalized probability in latent space according to the change of variables formula. During training we minimize the divergence between $\hat{f}(z)$ and the latent space distribution
\begin{align}
\begin{split}
    \loss^\text{inv}_F 
    &= D_F^z[\hat{f},\pl] 
     = \int \d z \,\pl(z) \: F\!\left(\frac{\hat{f}(z)}{\pl(z)}\right) \\
    &= \int \d z \,\pl(z) \: F\!\left(\frac{f(\gbar_\theta(z))}{\pl(z)} \left| \frac{\del \gbar_\theta(z)}{\del z} \right| \right) \\
    &= \int \d z \,\pl(z) \: F\left(\frac{f(\gbar_\theta(z))}{\overline{g}_\theta(z)} \right) 
    = \left\langle F\!\left(\frac{f(\gbar_\theta(z))}{\overline{g}_\theta(z)} \right) \right\rangle_{z \sim \pl(z)} \; .
\end{split}
\end{align}
We refer to this as inverse training because we evaluate the flow in the inverse direction. Note that this inverse training requires a differentiable integrand which is not needed in the forward training. It turns out that the forward and inverse training yield the same result for the loss,
\begin{align}
\begin{split}
    \loss^\text{inv}_F 
    &= \int \d x \,\pl(G_\theta(x)) \left| \frac{\del G_\theta(x)}{\del x} \right| \: F\!\left(\frac{f(\gbar_\theta(G_\theta(x)))}{\overline{g}_\theta(G_\theta(x))} \right) \\
    &= \int \d x \,g_\theta(x) \: F\!\left(\frac{f(x)}{g_\theta(x)} \right)
    = \loss^\text{fw}_F \; .
\end{split}
\end{align}
%

\subsubsection*{Loss gradients}

Because the two losses are identical, the gradients of the forward and inverse loss functions have the same expectation value,
\begin{align}
\begin{split}
    \nabla_\theta \loss^\text{inv}_F 
    =\left\langle \nabla_\theta F\!\left(\frac{f(\gbar_\theta(z))}{\overline{g}_\theta(z)} \right) \right\rangle_{z \sim \pl(z)}
    = \left\langle \nabla_\theta \frac{g_\theta(x)}{q(x)} \: F\!\left(\frac{f(x)}{g_\theta(x)} \right) \right\rangle_{x \sim q(x)}
    =\nabla_\theta \loss^\text{fw}_F \; .
\end{split}
\end{align}
From these expressions we can see that the two methods are not equivalent for the variances of the gradients. For the inverse training, it is 
\begin{align}
\begin{split}
    \text{Var}_{z\sim \pl(z)} \left( \nabla_\theta F\!\left(\frac{f(\gbar_\theta(z))}{\overline{g}_\theta(z)} \right) \right)
    &= \left\langle \left( \nabla_\theta F\!\left(\frac{f(\gbar_\theta(z))}{\overline{g}_\theta(z)} \right) - \nabla_\theta\loss^\text{inv}_F \right)^2 \right\rangle_{z\sim \pl(z)}\\
    &= \int \d z\, \pl(z) \left( \nabla_\theta F\!\left(\frac{f(\gbar_\theta(z))}{\overline{g}_\theta(z)} \right) - \nabla_\theta\loss^\text{inv}_F \right)^2 \; ,
\end{split}
\end{align}
while for the forward direction, we find 
\begin{align}
\begin{split}
    \text{Var}_{x\sim q(x)} \left( \nabla_\theta \frac{g_\theta(x)}{q(x)} \: F\!\left(\frac{f(x)}{g_\theta(x)} \right) \right)
    &= \left\langle \left( \nabla_\theta \frac{g_\theta(x)}{q(x)} \: F\!\left(\frac{f(x)}{g_\theta(x)} \right) - \nabla_\theta\loss^\text{fw}_F \right)^2 \right\rangle_{x\sim q(x)}\\
    &=\int \d x\,q(x) \left( \nabla_\theta \frac{g_\theta(x)}{q(x)} \: F\!\left(\frac{f(x)}{g_\theta(x)} \right) - \nabla_\theta\loss^\text{fw}_F \right)^2\; .
\end{split}
\end{align}
In general, the two will not be equal. Depending on the problem and choice of $F$-divergence, one of the two training modes can have noisier gradients, leading to a slower convergence of the training on finite batch sizes. In Appendix~\ref{app:analytic_loss_grads}, we study this behavior by analytically solving a simple 1D-toy example.

\subsection{Loss landscape}

\begin{figure}[b!]
    \includegraphics[width=0.33\linewidth]{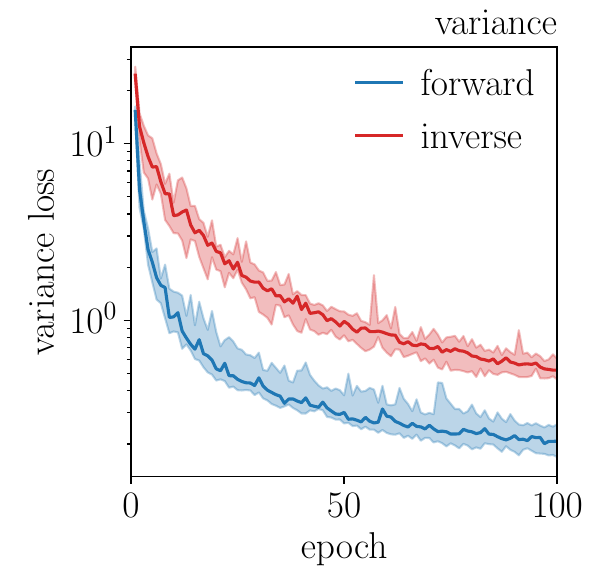}
    \includegraphics[width=0.33\linewidth]{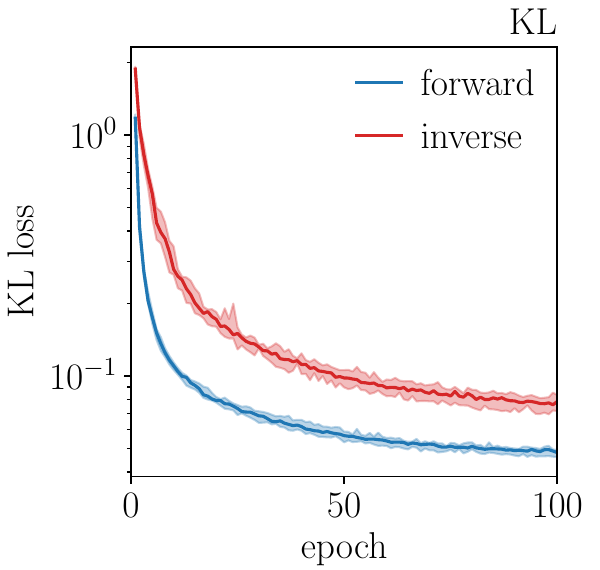}
    \includegraphics[width=0.33\linewidth]{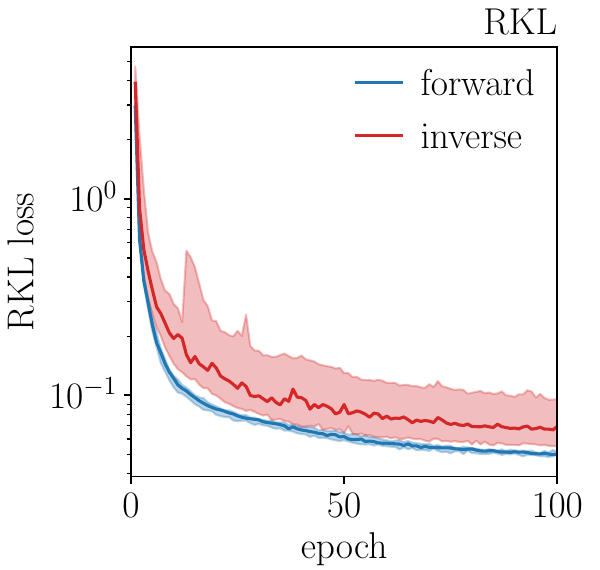}
    \caption{Loss values as a function of the training epochs for different losses: variance, KL-divergence, and RKL-divergence from left to right.}
    \label{fig:loss_lossfuncs}
\end{figure}

Finally, we look at established examples of divergences used to construct loss functions.
They use a normalized target distribution $f(x)$, which can be ensured during training by using batch-wise normalization of the integrand values:
\begin{itemize}
    \item variance, $F(t) = (t-1)^2$
    \begin{align}
        \loss^\text{fw}_\text{var} &= \left\langle \frac{g_\theta(x)}{q(x)} \: \left(\frac{f(x)}{g_\theta(x)} - 1\right)^2 \right\rangle_{x \sim q(x)} \notag \\
        \loss^\text{inv}_\text{var} &= \left\langle \left(\frac{f(\gbar_\theta(z))}{\overline{g}_\theta(z)} - 1 \right)^2 \right\rangle_{z \sim \pl(z)} \;  .
    \end{align}
    \item KL-divergence, $F(t) = t \log t$
    \begin{align}
        \loss^\text{fw}_\text{KL} &= \left\langle \frac{f(x)}{q(x)} \: \log \frac{f(x)}{g_\theta(x)} \right\rangle_{x \sim q(x)} \notag \\
        \loss^\text{inv}_\text{KL} &= \left\langle \frac{f(\gbar_\theta(z))}{\overline{g}_\theta(z)} \log \frac{f(\gbar_\theta(z))}{\overline{g}_\theta(z)} \right\rangle_{z \sim \pl(z)} \; .
    \end{align}
    \item reverse KL-divergence, $F(t) = -\log t$
    \begin{align}
        \loss^\text{fw}_\text{RKL} &= \left\langle \frac{g_\theta(x)}{q(x)} \: \log\frac{g_\theta(x)}{f(x)} \right\rangle_{x \sim q(x)} \notag \\
        \loss^\text{inv}_\text{RKL} &= \left\langle \log\frac{\overline{g}_\theta(z)}{f(\gbar_\theta(z))} \right\rangle_{z \sim \pl(z)} \; .
    \end{align}
\end{itemize}
We can test these six scenarios of forward and inverse training using three different divergences on triple-W production.
To this end, we run a simple single-channel \madnis training without buffered training, with the hyperparameters given in Tab.~\ref{tab:hyper_integrand}. 
We estimate the stability of the training and results by repeating each training ten times.
In Fig.~\ref{fig:loss_lossfuncs}, we first show the training behavior for the different scenarios. 
We immediately see that the KL-divergence leads to the most stable training. In contrast, the RKL-divergence combined with inverse training is the most noisy. In terms of performance measured by the final loss value, forward training consistently beats inverse training.

\begin{figure}[t!]
    \includegraphics[width=0.49\linewidth]{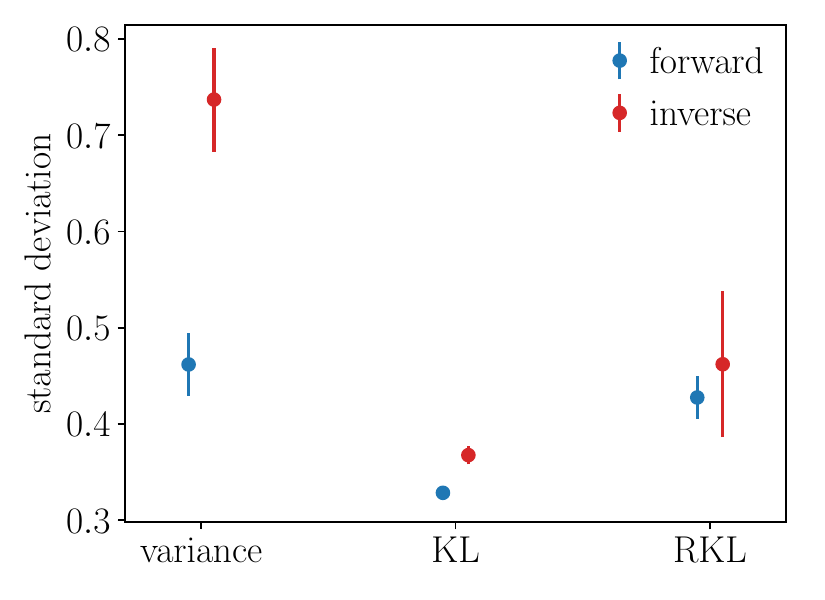}
    \includegraphics[width=0.49\linewidth]{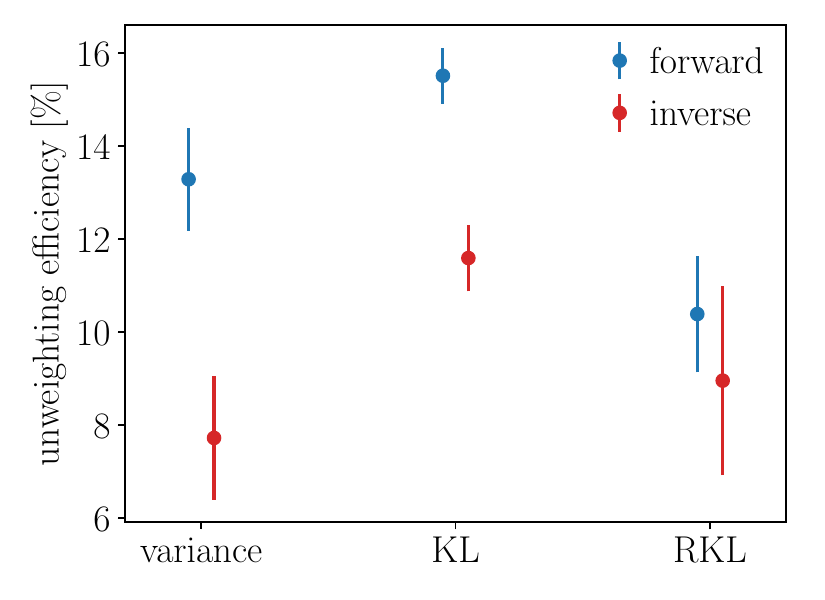}
    \caption{Relative standard deviations (left) and unweighting efficiencies (right) for different loss functions.}
    \label{fig:metrics_lossfuncs}
\end{figure}

In Fig.~\ref{fig:metrics_lossfuncs}, we show the unweighting efficiencies and the standard deviations of the integral as more quantitative quality measures.
For the standard deviation, we see that variance and RKL losses used for forward training lead to similar results, but the KL-divergence outperforms them. 
Also, in terms of unweighting efficiency, forward training with a KL-loss leads to the best results. 
The fact that RKL gives the worst unweighting efficiency is related to overweights, which the RKL does not penalize. 
This comparison has to be taken with a grain of salt, because the performance of forward training based on the variance and the KL-divergence are close in performance. An additional aspect we have to factor in is that a multi-channel loss can only be constructed using the variance, whereas the KL-divergence might be most suitable for single-channel integrals.

\subsubsection*{Additional derivatives}

\begin{figure}[b!]
    \includegraphics[width=0.49\linewidth]{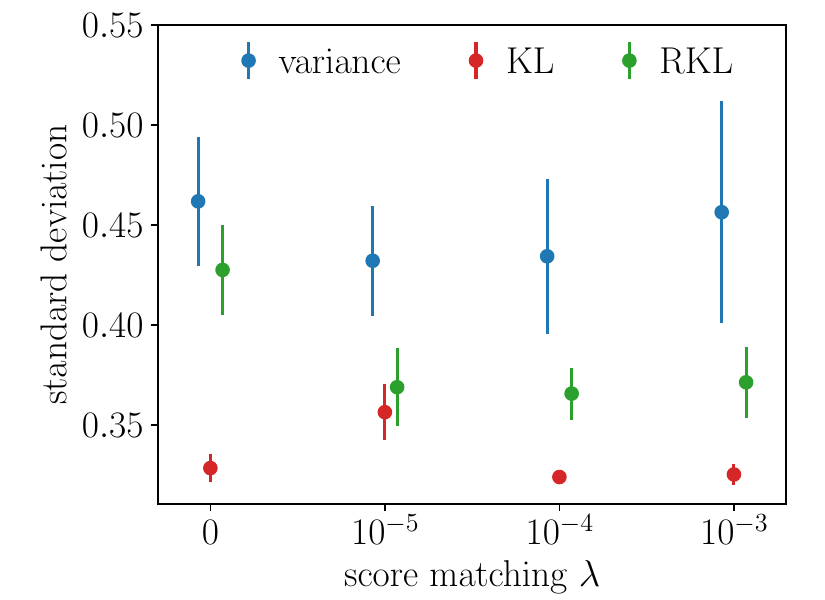}
    \includegraphics[width=0.49\linewidth]{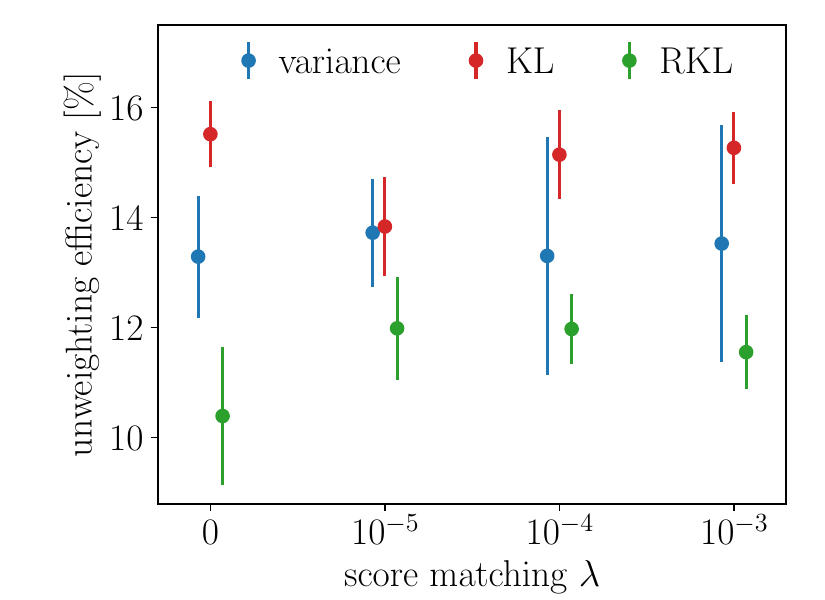}
    \caption{Relative standard deviations (left) and unweighting efficiencies (right) for different derivative matching coefficients.}
    \label{fig:metrics_score}
\end{figure}

When using differentiable integrands, we can also evaluate an additional 
derivative  matching term (also called score or force matching~\cite{Heinrich:2022xfa}) for each forward
loss introduced above,
\begin{align}
\loss^\text{fw} \to  
\loss^\text{fw} +  \lambda \left\langle\vert\partial_x\log f(x) - \partial_x \log g_{\theta}(x)\vert^2\right\rangle_{x\sim q(x)} \; .
\end{align}
The relative strength of the derivative term, $\lambda$, is a 
hyperparameter. In Fig.~\ref{fig:metrics_score}, we show the same triple-W
results as in Fig.~\ref{fig:metrics_lossfuncs}, but including derivative 
matching with different strengths $\lambda$.
For the variance and RKL losses, we see slight improvements in the results from 
the derivative matching. However, it turns out that it comes with less stable training. Altogether, the 
improvements from adding the derivative are relatively small, leading to the non-trivial question if their performance gain justifies the additional computational cost from evaluating the gradients. 
Moreover, generalizing this additional term to 
buffered training for multi-channel integration would 
come with a very large memory footprint.

\section{Differentiable phase space --- \madnis-Lite}
\label{sec:phase_space}

To define differentiable and trainable phase-space mappings, we re-introduce the relevant building blocks 
using consistent notation. For a hadron-collider process
\begin{align}
p_1 + p_2 \to k_1 + \cdots + k_n
\end{align}
the differential cross section is given by
\begin{align}
\begin{split}
\d\sigma &= \sum_{a,b}\d x_1\,\d x_2\,\d\Phi^{2\to n}\,f_a(x_1)f_b(x_2)\,\frac{(2\pi)^{4-3n}}{2x_1x_2s}\left\vert\mathcal{M}_{ab}(p_1,p_2|k_1,\dots,k_n)\right\vert^2\;,
\end{split}
\label{eq:diff_cross_section} 
\end{align}
with a sum over initial-state partons and the phase space density
\begin{align}
    \d\Phi^{2\to n} = \left[\prod_{i=1}^n \d^4 k_i \,\delta(k_i)\,\theta(k_i^0)\right]\,\delta^{(4)}\left(p_1+p_2-\sum_i k_i\right)\;.
\end{align}
The total cross section is 
\begin{align}
\begin{split}
    \sigma 
    = \int \d x_1\,\d x_2\,\d \Phi^{2\to n}(x)\,f(x) & \equiv \int \d x\,f(x) \\
    \mwith \quad 
    f(x) & =\frac{(2\pi)^{4-3n}}{2x_1x_2s}\sum_{a,b}\,f_a(x_1)f_b(x_2)\,\left\vert\mathcal{M}_{ab}(x)\right\vert^2 \; ,
    \label{eq:total_cross_section}
\end{split}
\end{align}
in terms of the phase space vector $x = (x_1,x_2, k_i)$.

\subsection{Parameterized mappings}
\label{sec:ps_mappings}

For a single channel, we choose a suitable mapping of  
$x$ to a unit-hypercube $z$,
\begin{align}
    \sigma &= \int \d x\,f(x) 
    = \int \d z\,\left.\frac{f(x)}{g(x)}\right|_{x=\gbar(z)}
\end{align}
To minimize the integration error, we choose the mapping $g$ to follow the peaking propagator structure of $f(x)$, as encoded in the Feynman diagrams. To this end, we decompose the phase-space integral into integrals over time-like invariants $s_i$, $2\to2$ scattering processes with $t$-channel propagators, and $1\to 2$ particle decays,
\begin{align}
\int \d \Phi^{2\to n}(x) =\prod_{i=1}^{n-2}\int\limits_{s_{i,\text{min}}}^{s_{i,\text{max}}}\d s_i\,
\prod_{j=1}^{\kappa}\int \d \Phi^{2\to 2}_j(x)\,
\prod_{k=1}^{n-\kappa-1}\int \d \Phi^{1\to 2}_k(x)\;.
\label{eq:ps_decomposition}
\end{align}
The number of $2\to2$ processes and $1\to 2$ decays depends on the diagram the mapping is based on. If several $t$-channel propagators are present, \ie for $\kappa \ge 2$, some of the $s_i$ do not correspond to propagators in the diagram and they are sampled uniformly. For the $(3n-4)$-dimensional integral this results in
\begin{itemize}
    \item $d_s= n-2$ degrees of freedom from time-like invariants;
    \item $d_{p}= 2\kappa$ degrees of freedom from $2\to 2$ scatterings.
    \item $d_d= 2(n-\kappa-1)$ degrees of freedom from decays;
\end{itemize}
Together with the PDF convolutions, this covers $3n-2$ integral dimensions. For each of these sub-integrals, it is possible to define appropriate mappings $x\leftrightarrow z$ as it is commonly done in many multi-purpose event generators~\cite{Sjostrand:2014zea, Alwall:2014hca, Sherpa:2019gpd}. We implement these physics-inspired mappings in a differentiable and invertible way using \pytorch, allowing us to perform forward training. In the following, we briefly review our parametrization~\cite{Dittmaier:2002ap} for each phase-space block needed to understand how they can be upgraded by additional trainable transformations.

\subsubsection*{Propagator invariants}

To construct a smooth mapping for a propagator 
\begin{align}
    |\mathcal{M}|^2\propto \frac{1}{(s-M^2)^2+M^2\Gamma^2} \;,
\end{align}
we map the invariant $s$ to a random number $z_s = G_\text{prop}(s)$ such that 
\begin{align}
\int\limits_{s_{\text{min}}}^{s_{\text{max}}}\d s 
= \int\limits_{0}^{1}
   \frac{\d z}{g_\text{prop} (s(z_s),s_{\text{min}},s_{\text{max}})} 
= \int\limits_{0}^{1}
   \d z \left\vert\frac{\partial \gbar_\text{prop}(z_s,m^2, s_{\text{min}},s_{\text{max}})}{\partial z}\right\vert \;.
\end{align}
Depending on the propagator width, we can introduce two different mappings. For a Breit-Wigner propagator we use 
\begin{align}
\begin{split}
    \gbar^\text{BW}_\text{prop}(z_s,m^2,s_{\text{min}},s_{\text{max}})&=m\Gamma\tan\left[y_1+(y_2-y_1)\,z_s\right]+m^2 \\
    g^\text{BW}_\text{prop}(s,m^2,s_{\text{min}},s_{\text{max}})&=\frac{m\Gamma}{(y_2-y_1)\left[(s-m^2)^2+m^2\Gamma^2\right]} \\
    y_{1/2} &=\arctan\left(\frac{s_{\text{min}/\text{max}}-m^2}{m\Gamma}\right) \;.
\end{split}
\label{eq:prop_massive}
\end{align}
For $\Gamma=0$ we instead employ
\begin{align}
\begin{split}
    \gbar^\nu_\text{prop}(z_s,m^2,s_{\text{min}},s_{\text{max}})&=\left[z_s(s_\text{max}-m^2)^{1-\nu}+(1-z_s)(s_\text{min}-m^2)^{1-\nu}\right]^{\frac{1}{1-\nu}} +m^2 \\
    g^\nu_\text{prop}(s,m^2,s_{\text{min}},s_{\text{max}})&=\frac{1-\nu}{\left[(s_\text{max}-m^2)^{1-\nu}-(s_\text{min}-m^2)^{1-\nu}\right](s-m^2)^\nu}\;,
    \label{eq:prop_stable_nu}
\end{split}
\end{align}
as long as $\nu\ne 1$. The parameter $\nu$ can be tuned, and the naive expectation $\nu=2$ is not 
necessarily the best choice. We choose $\nu=1.4$ and then optimize the mapping as explained below. 

\subsubsection*{$2\to 2$ scattering processes}

For a $2 \to 2$ scattering with $p_1 + p_2 = k_1 + k_2$, the momenta $p_{1,2}$ and the virtualities $k_i^2$ are fixed or sampled from other phase-space components. As this includes a $t$-channel propagator, we choose
\begin{align}
\begin{split}
\int \d \Phi^{2\to 2}(p_1,p_2;k^2_1,k^2_2)
&=\frac{1}{4\sqrt{\lambda(p^2,p_1^2,p_2^2)}}\int\limits_0^{2\pi}\d\phi^*\int\limits_{-t_\text{max}}^{-t_\text{min}}\d |t| \\
&=\int\limits_{0}^{1}\frac{\d z_\phi \d z_t}{g_{2\to 2}(p^2,p_1^2,p_2^2,t,m^2,\nu,t_\text{min},t_\text{max})}\; ,
\end{split}
\end{align} 
where $\phi^*$ is the azimuthal angle defined by $p_1$ and $k_1$ in the CM frame of $p=p_1+p_2$, and $\lambda(x,y,z)=(x-y-z)^2-4yz$. The invariant $t=(p_1-k_1)^2<0$ depends only linearly on the azimuthal angle $\cos\theta^*$ as
\begin{align}
t=k_1^2+p_1^2-\frac{(p^2+k_1^2-k_2^2)(p^2+p_1^2-p_2^2)-\sqrt{\lambda(p^2,k_1^2,k_2^2)}\sqrt{\lambda(p^2,p_1^2,p_2^2)}\cos\theta^*}{2p^2}\;.
\label{eq:t_invariant_theta}
\end{align}
The integration boundaries can be calculated from this with $-1\le \cos\theta^*\le 1$. We sample the polar angle and $t$ according to
\begin{align}
    \phi^*=2\pi z_\phi \quad \mand \quad |t|=\gbar^\nu_\text{prop}(z_t,m^2,-t_{\text{max}},-t_{\text{min}}) \; ,
    \label{eq:scattering_sampling}
\end{align}
with, correspondingly,
\begin{align}
g_{2\to 2}(p^2,p_1^2,p_2^2,t,m^2,\nu,t_\text{min},t_\text{max})=\frac{2}{\pi}\sqrt{\lambda(p^2,p_1^2,p_2^2)}\,g_\text{prop}(-t,m^2,\nu,-t_{\text{max}},-t_{\text{min}})\;.
\end{align}
Further, for each $t$-channel block in our diagram, the corresponding $s$-invariant, \ie $p^2=s$, also needs to be sampled as time-like invariant in  Eq.\eqref{eq:ps_decomposition}. However, in contrast to invariants reflecting propagators in the diagram, these invariants belong to pseudo particles and can be sampled flat
\begin{align}
\begin{split}
    \gbar_\text{pseudo}(z_s,s_{\text{min}},s_{\text{max}})&=z_s\,(s_\text{max}-s_\text{min})+s_\text{min} \\
    g_\text{pseudo}(s,s_{\text{min}},s_{\text{max}})&=\frac{1}{s_\text{max}-s_\text{min}}\;.
    \label{eq:flat_invariants}
\end{split}
\end{align}
%

\subsubsection*{$1\to 2$ particle decays}

For isotropic decays with $p=k_1+k_2$, the momentum $p$ and the virtualities $k_i^2$ are again sampled from other phase-space components. We choose the 
polar angle $\phi^*$ and azimuthal angle $\theta^*$ in the decay rest frame 
as integration variables and sample $\phi^*=2\pi z_\phi$ and $\cos\theta^*=2 z_\theta - 1$ uniformly,
\begin{align}
\begin{split}
\int \d \Phi^{1\to 2}(p;k^2_1,k^2_2)
=\frac{\sqrt{\lambda(p^2,k_1^2,k_2^2)}}{8p^2}\int\limits_0^{2\pi}\d\phi^*\int\limits_{-1}^{1}\d\cos\theta^* 
=\frac{1}{g_\text{decay}(p^2,k^2_1,k^2_2)}\int\limits_{0}^{1}\d z_1 \d z_2 \\
\mwith \qquad g_\text{decay}(p^2,k^2_1,k^2_2)=\frac{2p^2}{\pi\sqrt{\lambda(p^2,k_1^2,k_2^2)}}\;.
\end{split}
\label{eq:sample_decay}
\end{align}
%

\subsubsection*{PDF convolutions}

For the PDF convolutions, we introduce $\tau = x_1x_2$, such that
the squared partonic CM energy is given by $\hat{s}=\tau s$. This allows us to write 
\begin{align}
\begin{split}
    \int\limits_{0}^{1} \d x_1 \d x_2\,\Theta(\hat{s}-\hat{s}_\text{min})
    = \int\limits^{1}_{\tau_\text{min}} \d\tau
    \int\limits_{\tau}^{1} \frac{\d x_1}{x_1} 
    =\int\limits_{0}^{1} \frac{\d z_\tau \d z_{x_1}}{g_\text{lumi}(\tau, \tau_\text{min})} \\
    \mwith \qquad g_\text{lumi}(\tau, \tau_\text{min})=\frac{1}{\tau\,\log\tau\,\log\tau_\text{min}}\;,
\end{split}
\end{align}
where $\hat{s}_\text{min}$ follows from final-state masses and cuts and we sample
\begin{align}
    \tau=\tau_\text{min}^{1-z_\tau} \qquad \mand \qquad x_1=\tau^{z_{x_1}} \; .
    \label{eq:lumi_sampling}
\end{align}
The induced density $g_\text{lumi}$ exactly cancels the flux factor $\tau^{-1}$ in Eq.\eqref{eq:diff_cross_section}. If there are no $t$-channels, \ie $\kappa=0$, the squared CM energy $\hat{s}$ also belongs to a propagator in the diagram. In this case, it is beneficial to sample $\tau$ such that this propagator structure is mapped out.\medskip

Each of the $s$-invariants, $2\to 2$ scatterings, and decay blocks described above transform one or two random numbers. 
They can appear multiple times for a given Feynman diagram, as illustrated in Fig.~\ref{fig:parametrization}.
In Appendix~\ref{app:channe_mapping_example}, we illustrate how these components are combined to parametrize a complete channel mapping for $\PW+4~\jets$ production.

\subsection{Learnable bilinear spline flows}
\label{sec:bilinear_flow}

\begin{figure}[b!]
    \centering
    \includegraphics[width=0.7\linewidth]{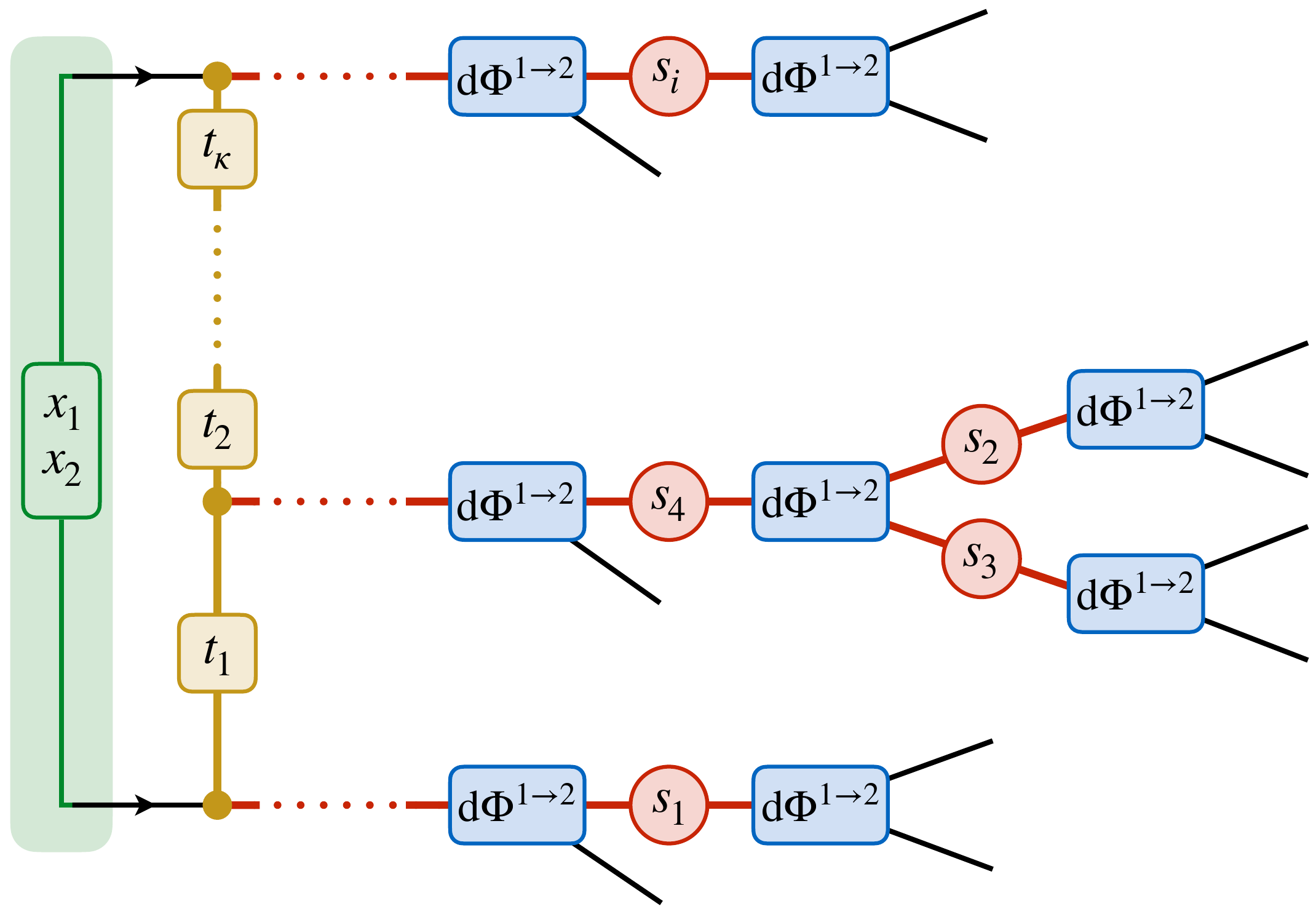}
    \caption{Topological diagram illustrating our separable and differentiable phase-space mappings. Each colored block represents one of the introduced components which can be modified by a trainable bilinear flow. }
    \label{fig:parametrization}
\end{figure}

For a typical \madnis training, the flow sub-networks often encode relatively simple functions. 
For these cases, we introduce bilinear spline flows to replace the sub-networks with second-order polynomials. 
A $d_x$-dimensional transformation $x \leftrightarrow z$ with a $d_c$-dimensional condition $c$ can be written as
\begin{align}
    z = G(x; W \hat{c}) \qquad\mwith\qquad
    \hat{c} = 
    \begin{pmatrix}
        1 \\ c_i \\ c_i c_j 
    \end{pmatrix}  
    \quad \text{for} \quad i \leq j \; ,
\end{align}
where $G$ is a rational quadratic spline transformation and $W$ is a trainable matrix. The number of trainable parameters for such a transformation with $n_b$ bins is
\begin{align}
    d_W = (3 n_b + 1) \times d_x \times \left( 1 + d_c + \frac{1}{2} d_c (d_c+1) \right) \; .
\end{align}
This way, we can build small and fast, but sufficiently expressive trainable transformations for a small number of dimensions $d_x$ and $d_c$. 
Another benefit is the interpretability of bilinear spline flows because $W$ tells us how 
strongly the spline transformation is correlated with the conditional inputs.

We can combine these trainable mappings with the propagator, decay, scattering, and PDF blocks introduced above and use them to transform their uniform random number input. For mappings with two random numbers, we allow for correlations between the two dimensions. Because all parts of the phase-space mappings are differentiable, the bilinear flow can even be conditional on intermediate physical features that are available only during the evaluation of the phase-space mapping. This enhances the expressivity and interpretability of the learned transformation. 

\begin{table}[b!]
    \centering
    \begin{small} \begin{tabular}[t]{lll}
    \toprule
    Mapping & Parameters & Conditions\\
    \midrule
    Time-like invariants, Eqs.\eqref{eq:prop_massive},\eqref{eq:prop_stable_nu} & 190 & partonic CM energy $\sqrt{\hat{s}/s_\text{lab}}$ \\
    (separate for massless and & & minimal decay CM energy $\sqrt{s_\text{min}/s_\text{lab}}$ \\
    massive propagators) & & maximal decay CM energy $\sqrt{s_\text{max}/s_\text{lab}}$ \\
    \midrule
    $2\to 2$ scattering, Eq.\eqref{eq:scattering_sampling} & 798 & correlations between $z_t$, $z_\phi$ \\
    & & partonic CM energy $\sqrt{\hat{s}/s_\text{lab}}$ \\
    & & scattering CM energy $\sqrt{p^2/s_\text{lab}}$ \\
    & & virtualities $\sqrt{k_{1,2}^2 / s_\text{lab}}$ \\
    \midrule
    Time-like invariants for& 190 & partonic CM energy $\sqrt{\hat{s}/s_\text{lab}}$ \\
    pseudo-particles, Eq.\eqref{eq:flat_invariants}& & minimal energy $\sqrt{s_\text{min}/s_\text{lab}}$ \\
    & & maximal energy $\sqrt{s_\text{max}/s_\text{lab}}$ \\
    \midrule
    $1\to 2$ decay, Eq.\eqref{eq:sample_decay} & 380 & correlations between $z_\theta$, $z_\phi$ \\
    & & partonic CM energy $\sqrt{\hat{s}/s_\text{lab}}$ \\
    & & decay CM energy $\sqrt{p^2/s_\text{lab}}$ \\
    \midrule
    PDF convolutions, Eq.\eqref{eq:lumi_sampling} & 114 & correlations between $z_\tau$, $z_{x_1}$\\
    \bottomrule
    \end{tabular} \end{small}
    \caption{Trainable components}
    \label{tab:trainable_components}
\end{table}

\subsubsection*{Implementation}

We implement the trainable bilinear spline flows with 6 spline bins. We list the trainable components of the phase space mappings, the conditional features, and the number of trainable parameters in Tab.~\ref{tab:trainable_components}. 
These parameters are shared between channels and multiple instances of the same block in one channel. This way, the number of trainable parameters stays the same for different processes and allows the use of mappings trained on one process, like $\PW+3~\jets$, for reasonably related other processes. Note that for processes like  $\PW+4~\jets$ ($\Pt\Ptbar+3~\jets$) with up to $384$ ($945$) integration channels, this parameter sharing reduces the computational cost significantly.

We train \madnis-Lite using the multi-channel variance loss from Eq.\eqref{eq:madnis_loss}, but without trainable channel weights. We use stratified training to focus the available training samples on channels with a large contribution to the total cross section, and buffered training to reduce the number of integrand evaluations. The training hyperparameters are given in Tab.~\ref{tab:hyper_phase_space}.

\subsubsection*{Performance}

\begin{figure}[t]
    \includegraphics[width=0.49\linewidth]{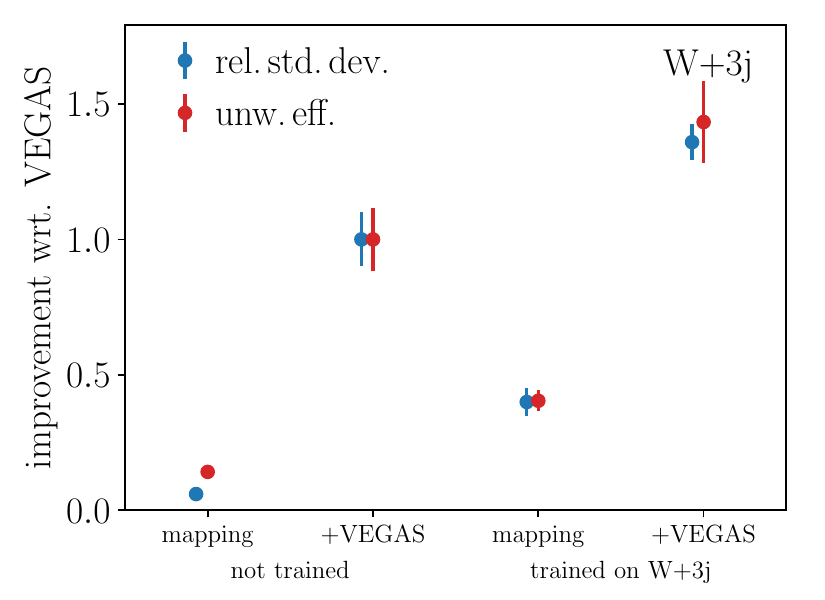}
    \includegraphics[width=0.49\linewidth]{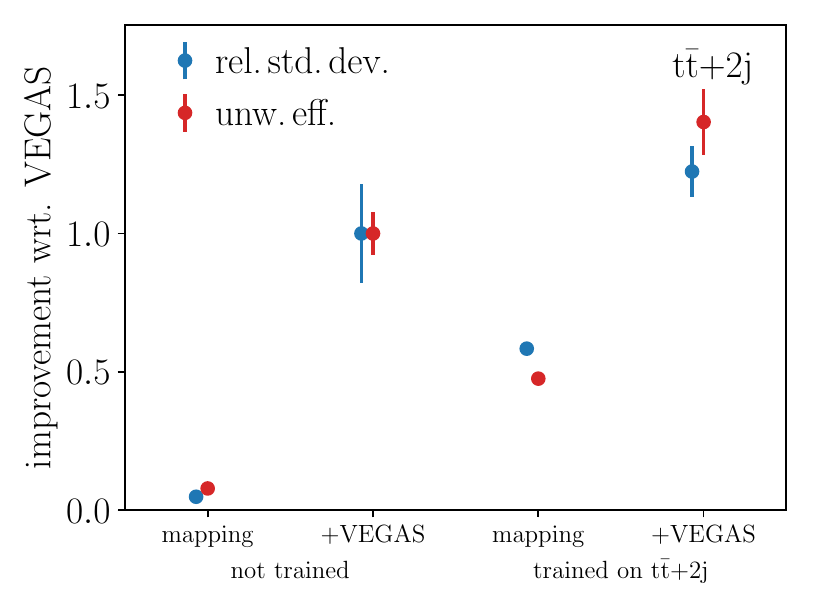}\\
    \includegraphics[width=0.49\linewidth]{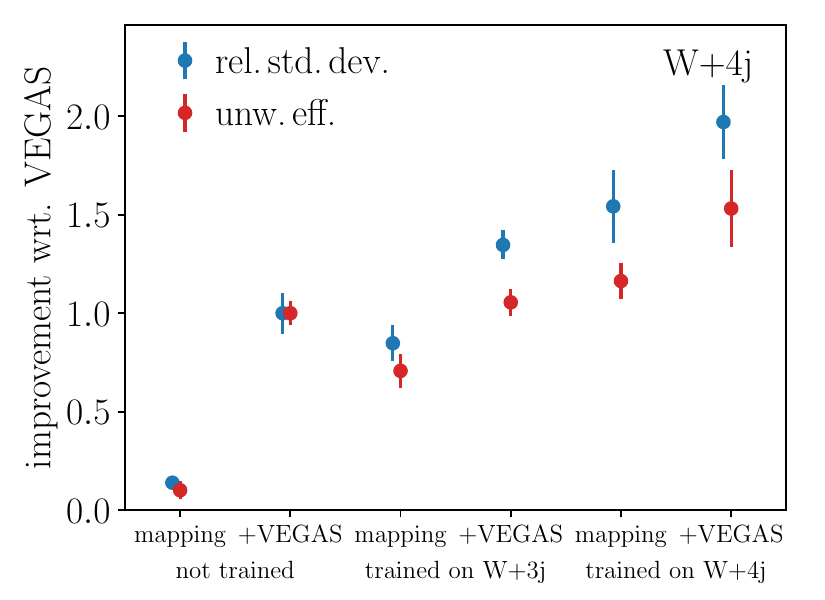}
    \includegraphics[width=0.49\linewidth]{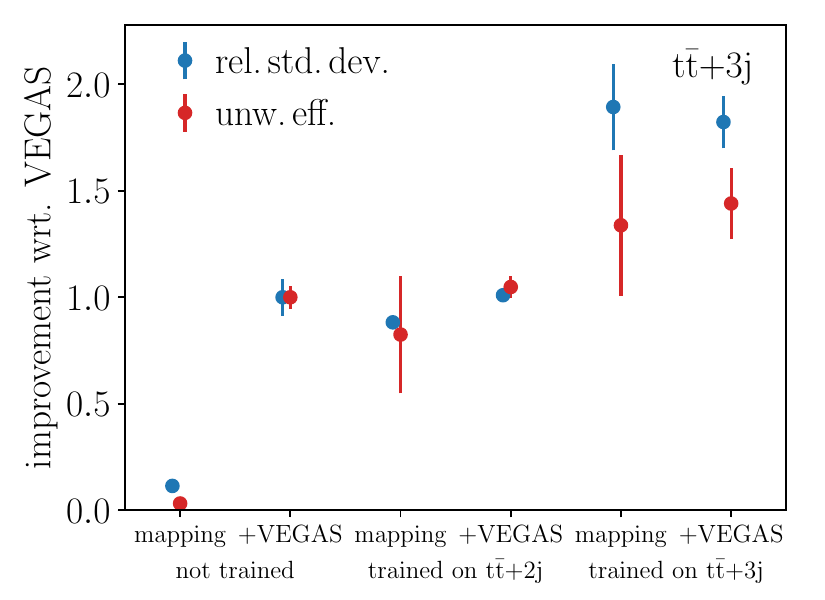}
    \caption{Improvement of the unweighting efficiency and relative standard deviation of different setups with respect to an untrained phase space mapping refined with VEGAS for W+jets (left) and $\text{t}\bar{\text{t}}$+jets (right).}
    \label{fig:metrics_phasespace}
\end{figure}

In Fig.~\ref{fig:metrics_phasespace}, we compare the unweighting efficiencies and relative integration errors for different scenarios and different processes. Throughout all considered processes, we benchmark our trained mappings against the raw mappings combined with and without \vegas. The shown error bars were obtained by running the integration, including \vegas optimization if applicable, ten times and taking the mean and standard deviation. All results are shown relative to the \vegas performance without trained mappings.

We start by considering the $\PW+3~\jets$ process in the upper left plot of Fig.~\ref{fig:metrics_phasespace}. The trained mappings without additional \vegas optimization outperform the raw phase-space mapping but are a bit worse than the \vegas optimized mappings. This is because the number of trainable parameters of our bilinear flow is quite small as it is shared among multiple channels and building blocks, see Tab.~\ref{tab:trainable_components}. In contrast, \vegas builds an independent grid for each channel and phase-space direction, resulting in more than 30k optimized parameters. When combining our trained mapping with \vegas, we achieve the best performance in the $\PW+3~\jets$ scenario, with an improvement factor of up to $1.5$. For the $\Pt\Ptbar+2~\jets$ scenario in the upper right plot, the story is the same.

Next, we consider the same processes but with an additional jet in the final state. The results for different scenarios are shown in the lower two plots in Fig.~\ref{fig:metrics_phasespace}. Again, we consider the mapping that has been trained on the $\PW+3~\jets$ process and evaluate it on the $\PW+4~\jets$ process without further training. We find that the pre-trained mappings are very close in performance to the \vegas benchmark, without any specific optimization on the $\PW+4~\jets$ process. Like before, when additionally combining with \vegas we outperform our untrained phase-space mappings. If we directly train our mappings on $\PW+4~\jets$, we immediately outperform our untrained benchmark mappings even without further optimizing with \vegas. When combining the trained mappings with an additional \vegas optimization, we achieve an improvement factor of up to 2 for the $\PW+4~\jets$ process. 

Again, when turning to the $\Pt\Ptbar+3~\jets$ scenario, we observe the same behavior. This indicates that our trainable mappings work well and are capable of generalizing from one process to another process with an additional final state jet. This means our trainable bilinear flow represents the smallest foundation model possible. We note that going even one step further by pre-training our bilinear flows on $\PW+2~\jets$ and $\Pt\Ptbar+1~\jets$, respectively, does not generalize well to higher multiplicities as these low-multiplicity processes are too simple to encode all the necessary information.

\subsubsection*{Explainability}

\begin{figure}[t]
    \includegraphics[width=0.49\linewidth,page=9]{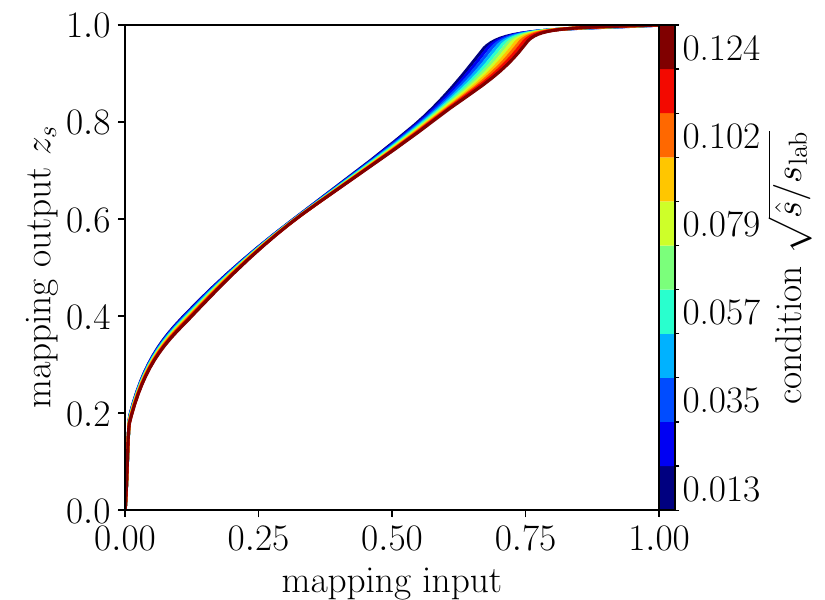}
    \includegraphics[width=0.49\linewidth,page=19]{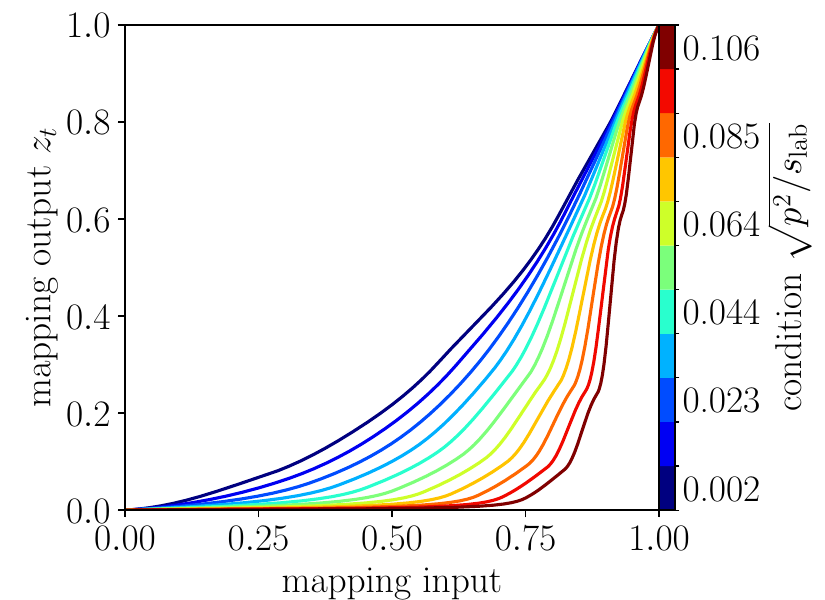}
    \caption{Mappings learned by the bilinear spline flow for W+3~jets. Left panel: Learned mapping for the time-like invariant for massless propagators, conditional on the partonic CM energy $\sqrt{\hat{s}}$. Right panel: Learned mapping for the $t$-invariant in $2\to 2$ scatterings, conditional on the scattering CM energy $\sqrt{p^2}$.}
    \label{fig:rainbow}
\end{figure}

Another benefit of using our bilinear-flow-enhanced mappings is the possibility to understand and interpret the learned correlations. As an example, we consider the learned transformation for the $\PW+3~\jets$ process in Fig.~\ref{fig:rainbow}. Both plots show a learned transformation of an input of one of the phase-space blocks conditioned on some physical features relevant to that component. In the left panel of Fig.~\ref{fig:rainbow}, we consider the learned transformation for a massless propagator conditional on the partonic CM energy $\sqrt{\hat{s}}$. We can see that the overall shape of the mapping deviates from the flat mapping, being slightly bulged upwards. This means that our fixed choice of $\nu=1.4$ was slightly too large, indicating stronger pole cancellations in the collinear limit. Further, the mapping tends to avoid $z_s<0.2$ and hence avoiding to sample the $s_\text{min}$ region due to $p_\text{T}$ cuts on the final state jets. In contrast, the mapping favors sampling into $s\approx s_\text{max}$ stemming from momentum conservation in dominating integration channels containing only $s$-channels.
This means, we possibly need to allow for a more flexible optimization of the time-like invariants depending on the underlying topology or the linked particle id. On top of this overall correlation, we can also look into the dependence on the $\sqrt{\hat{s}}$. The condition is varied between $2.5\%$ and $97.5\%$ of the quantile from the distribution of values that this block sees during event generation. We can observe that varying $\sqrt{\hat{s}}$ only has a very small effect on the mapping of the $s$-invariant, indicating a small correlation.

For the right panel of Fig.~\ref{fig:rainbow}, we consider the learned mapping for the $t$-invariant in a $2\to 2$ scattering block conditioned on the CM energy $\sqrt{p^2}$ of that $2\to 2$ scattering. In this case, varying $\sqrt{p^2}$ has a large influence on the optimal $t$-invariant mapping. This means that for larger $p^2$, the mapping tends to sample smaller $z_t$, which is physically linked to smaller scattering angles $\theta^*$ in the center-of-mass system of the $2\to 2$ scattering block and thus prefers very forward scattering.

\section{Outlook}

Modern machine learning is a promising path to improve the critical multi-purpose event generators to the speed and precision level required by the HL-LHC. For \textsc{MadGraph}, the \madnis~\cite{Heimel:2022wyj,Heimel:2023ngj} project has shown that significant gains can be realized by implementing multi-channel importance sampling using modern neural networks.

An exciting and equally promising new approach is differentiable programming applied to event generation~\cite{Heinrich:2022xfa}. We have tested the potential gains from a differentiable \madnis for two setups. First, we have developed a fully differentiable combination of matrix element, phase space, and parton densities to use derivative information for optimal integration and sampling. While this setup works well, we have not found significant improvements over the established \madnis methodology. 

As a second and more lightweight use of differentiable code, we have developed a new, modular, and differentiable phase-space mapping. This \madnis-Lite approach factorizes the phase space into differentiable standard blocks, each consisting of a physics-inspired mapping and a learnable bilinear spline flow. These blocks have the great advantage of being economical and generalizable, so they are much easier to train and use than the full \madnis framework. For a set of benchmark processes, they show great promise and low computational cost relative to the full \madnis.

Altogether, this allows us to combine three strategies for future \textsc{MadGraph} releases: (i) the standard multi-channel importance sampling using \vegas techniques provides fast results whenever the integrand factorizes at least approximately; (ii) for more complex Feynman diagrams with factorizing physics structures, which might not be easily mapped on individual phase space directions, \madnis-Lite provides efficient and fast ML-integration and sampling; (iii) for the highest precision and general matrix elements, including, for instance, gauge cancellations between Feynman diagrams, \madnis leads to significant improvements over established methods. In combination, these methods provide the ML backbone for optimal event generation at the HL-LHC.

\subsection*{Acknowledgements}
First, we would like to thank Michael Kagan and Lukas Heinrich for continuous encouragement by asking all the right questions many times.
OM and RW acknowledge support by FRS-FNRS (Belgian National Scientific Research Fund) IISN projects  4.4503.16. TP and TH are supported by the Deutsche
Forschungsgemeinschaft (DFG, German Research Foundation) under grant
396021762 -- TRR~257 \textsl{Particle Physics Phenomenology after the
Higgs Discovery}. TH is funded by the Carl-Zeiss-Stiftung through the
project \textsl{Model-Based AI: Physical Models and Deep Learning for
Imaging and Cancer Treatment}. This research is supported by the
Deutsche Forschungsgemeinschaft (DFG, German Research Foundation)
through Germany's
Excellence Strategy EXC~2181/1 -- 390900948 (the \textsl{Heidelberg
  STRUCTURES Excellence Cluster}). 
The authors acknowledge support by the state of Baden-Württemberg through bwHPC and the German Research Foundation (DFG) through grant no INST 39/963-1 FUGG (bwForCluster NEMO).

\clearpage
\appendix
\section{Hyperparameters}
\label{app:hyperparams}

\begin{table}[h!]
    \centering
    \begin{small} \begin{tabular}[t]{l|l}
    \toprule
    Parameter & Value \\
    \midrule
    Optimizer & Adam~\cite{Kingma:2014vow} \\
    Learning rate & 0.001 \\
    LR schedule & exponential \\
    Final learning rate & 0.0001 \\
    Batch size & 1024 \\
    Training length & 10k batches \\
    Permutations & Logarithmic decomposition~\cite{Gao:2020vdv} \\
    Number of coupling blocks & $2 \: \lceil \log_2 D \rceil=6$ \\
    Coupling transformation & RQ splines~\cite{durkan2019neural} \\
    Subnet hidden nodes & 32 \\
    Subnet depth & 3 \\
    Activation function & leaky ReLU \\
    \bottomrule
    \end{tabular} \end{small}
    \caption{\madnis hyperparameters used in Sec.~\ref{sec:integrand}.}
    \label{tab:hyper_integrand}
\end{table}

\begin{table}[h!]
    \centering
    \begin{small} \begin{tabular}[t]{l|l}
    \toprule
    Parameter & Value \\
    \midrule
    Optimizer & Adam \\
    Learning rate & 0.01 \\
    LR schedule & exponential \\
    Final learning rate & 0.001 \\
    Batch size & $\min(200 \cdot n_c^{0.8}, 10000)$ \\
    Buffered training gain~\cite{Heimel:2022wyj} & $6$ \\
    Training length & 7.8k batches \\
    Uniform training fraction ~\cite{Heimel:2023ngj} & 0.1 \\
    RQ spline bins & 6 \\
    \vegas iterations & 7\\
    \vegas bins & 64 \\
    \vegas samples per iteration & 20k\\
    \vegas damping $\alpha$ & 0.7 \\
    \bottomrule
    \end{tabular} \end{small}
    \caption{\madnis-Lite and \vegas hyperparameters used in Sec.~\ref{sec:phase_space}.}
    \label{tab:hyper_phase_space}
\end{table}

\section{Analytic loss function gradients}
\label{app:analytic_loss_grads}

\begin{figure}[b!]
    \includegraphics[width=0.33\linewidth]{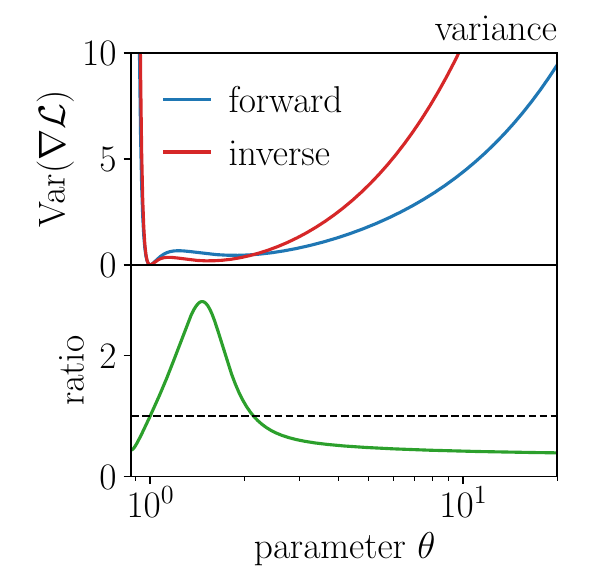}
    \includegraphics[width=0.33\linewidth]{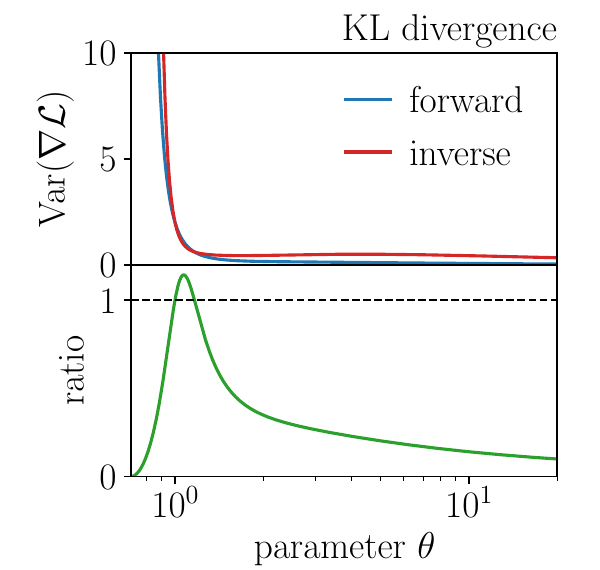}
    \includegraphics[width=0.33\linewidth]{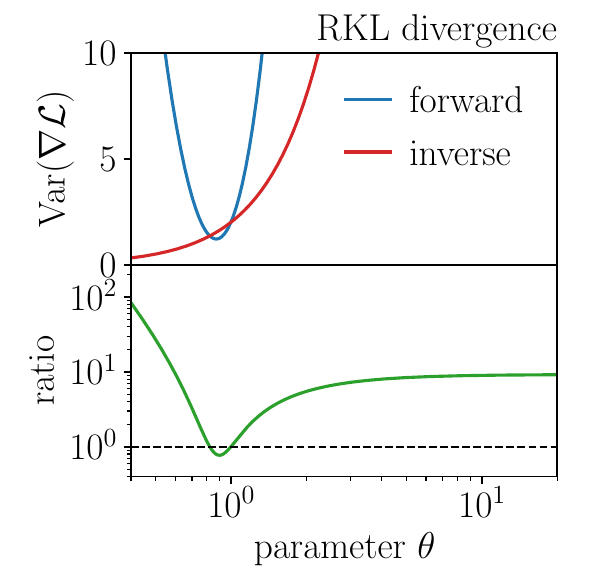}
    \caption{Analytic solutions for the variance of the gradients of different loss functions}
    \label{fig:analytic_loss}
\end{figure}

To exemplify the properties of the forward and inverse loss, we consider a simple 1-dimensional target function $f$ as
\begin{align}
    f(x)=\frac{1}{\sqrt{2\pi}}\exp\left(-\frac{x^2}{2}\right)\;.
\end{align}
The latent distribution is given as a standard normal distribution as
\begin{align}
    \pl(z)=\frac{1}{\sqrt{2\pi}}\exp\left(-\frac{z^2}{2}\right)\;.
\end{align}
We now assume a simple trainable mapping (1D flow if you want), given by
\begin{align}
    x\equiv \overline{G}_\theta(z)=\theta\cdot z\quad\longleftrightarrow \quad z\equiv G_\theta(x)=\frac{z}{\theta}\;.
\end{align}
inducing the Jacobian determinants
\begin{align}
    \left\vert\frac{\partial G_\theta(x)}{\partial x}\right\vert=\frac{1}{\theta} \quad \longleftrightarrow \quad
    \left\vert\frac{\partial \overline{G}_\theta(z)}{\partial z}\right\vert=\theta\;.
\end{align}
Combining this with the prior sample density yields the overall importance sampling (pseudo) density
\begin{align}
    g_\theta(x)&=\pl(G_\theta(x))\left\vert\frac{\partial G_\theta(x)}{\partial x}\right\vert=\frac{1}{\sqrt{2\pi \theta^2}}\exp\left(-\frac{x^2}{2\theta^2}\right)\\[0.2cm]
    \overline{g}_\theta(z)&=\pl(z)\left\vert\frac{\partial \overline{G}_\theta(z)}{\partial z}\right\vert^{-1}=\frac{1}{\sqrt{2\pi \theta^2}}\exp\left(-\frac{z^2}{2}\right)\qquad \mwith\quad \overline{g}_\theta(G_\theta(x)) = g_\theta(x) \;.
\end{align}
In this case, the integral of $f$ can be calculated as
\begin{align}
    I =\int\d x\;f(x)
    =\int\d x\;g_\theta(x)\,\frac{f(x)}{g_\theta(x)}
    =\left\langle\frac{g_\theta(x)}{q(x)}\,\frac{f(x)}{g_\theta(x)}\right\rangle_{x\sim q(x)}=1\;.
\end{align}
%

\subsubsection*{Variance loss}
While the integral does not change, the variance of the integrand is given by
\begin{align}
\begin{split}
    \loss^\text{fw}_\text{inv}=\loss^\text{inv}_\text{var}&=\left\langle\frac{g_\theta(x)}{q(x)}\,\left(\frac{f(x)} {g_\theta(x)}-1\right)^2\right\rangle_{x\sim q(x)} \\
    &=\int\d x\,g_\theta(x)\,\left(\frac{f(x)}{g_\theta(x)}-1\right)^2  \\
    &=\frac{\theta^2}{\sqrt{2\theta^2-1}}-1 \qquad \mfor \quad \theta > \frac{1}{\sqrt{2}}\;,
\end{split}
\end{align}
which is exactly zero for the expected value of $\theta=1$. Next, we can calculate the expectation value of the gradient of the integrand with respect to $\theta$, as this quantity is used during optimization. 
This yields
\begin{align}
\begin{split}
\nabla_\theta \loss^\text{fw}_\text{var}=\nabla_\theta \loss^\text{inv}_\text{var}
&=\frac{2\theta\,(\theta^2-1)}{(2 \theta^2-1)^{3/2}} \; .
\end{split}
\end{align}
On the other hand, the variance of the gradient is given by
\begin{align}
\begin{split}
\text{Var}_{x\sim q(x)}
&=\left\langle\left(\nabla_\theta\frac{g_\theta(x)}{q(x)}\,\left(\frac{f(x)}{g_\theta(x)}-1\right)^2-\nabla_\theta \loss^\text{fw}_\text{var}\right)^2\right\rangle_{x\sim q(x)}\\
&=\int \d x \,q(x)\left(\nabla_\theta\frac{g_\theta(x)}{q(x)}\,\left(\frac{f(x)}{g_\theta(x)}-1\right)^2-\nabla_\theta \loss^\text{fw}_\text{var}\right)^2\\
&=-\frac{1}{4}\left(2-\frac{8}{\theta^2}-\frac{2}{(2\theta^2-1)^2}+\frac{2}{(2\theta^2-1)^3}+\frac{24}{(2\theta^2-1)^{5/2}}\right.\\
&-\frac{16}{(2\theta^2-1)^{3/2}}-\frac{2}{2\theta^2-1}+\frac{8}{(2\theta^2-1)^{1/2}}\\
&-\frac{9}{(4\theta^2-3)^{5/2}}+\frac{3}{(4\theta^2-3)^{3/2}}
\left.-\frac{1}{(4\theta^2-3)^{1/2}}-\sqrt{4\theta^2-3} \right) \; .
\end{split}
\end{align}
However, for the inverse training, we obtain the variance of the gradient as
\begin{align}
\begin{split}
\text{Var}_{z\sim \pl(z)}
&=\left\langle\left(\nabla_\theta\left(\frac{f(\overline{G}_\theta(z))}{\overline{g}_\theta(z)}-1\right)^2-\nabla_\theta \loss^\text{inv}_\text{var}\right)^2\right\rangle_{z\sim \pl(z)}\\
&=\int \d z \,\pl(z)\left(\nabla_\theta\left(\frac{f(\overline{G}_\theta(z))}{\overline{g}_\theta(z)}-1\right)^2-\nabla_\theta \loss^\text{inv}_\text{var}\right)^2\\
&=-\frac{1}{16}\left(8-\frac{8}{(2\theta^2-1)^2}+\frac{8}{(2\theta^2-1)^3}-\frac{48}{(2\theta^2-1)^{5/2}}\right.\\
&-\frac{32}{(2\theta^2-1)^{3/2}}-\frac{8}{2\theta^2-1}-\frac{48}{(2\theta^2-1)^{1/2}}\\
&-\frac{81}{(4\theta^2-3)^{5/2}}-\frac{9}{(4\theta^2-3)^{3/2}}-\frac{27}{(4\theta^2-3)^{1/2}}\\
&\left.-11\sqrt{4\theta^2-3}+\frac{256\theta\,(3 \theta^4- 4 \theta^2+2)}{(3\theta^2-2)^{5/2}} \right) \; .
\end{split}
\end{align}
%

\subsubsection*{KL loss}

For the expectation value of the gradient, we obtain for both directions
\begin{align}
    \nabla_\theta \loss_{\text{KL}}^\text{fw}
    =\nabla_\theta \loss_{\text{KL}}^\text{inv}
    =\frac{\theta^2-1}{\theta^3}\;.
\end{align}
While for the variance, we obtain
\begin{align}
    \text{Var}\left(\nabla_\theta \loss_{\text{KL}}^\text{fw}\right)
    &=-\frac{1}{\theta^6}+\frac{2}{\theta^4}-\frac{1}{\theta^2} + \frac{4 \theta^4 - 8 \theta^2 + 6}{(2\theta^2-1)^{5/2}}\\
    \begin{split}
    \text{Var}\left(\nabla_\theta \loss_{\text{KL}}^\text{inv}\right)
    &=\frac{-1}{4\theta^6(2\theta^2-1)^{9/2}}\left(
    \sqrt{2\theta^2-1}\,\left(4-40\theta^2+164\theta^4\right)\right.\\
    &-\theta^6\left[3+49\theta^8+352\sqrt{2\theta^2-1}-4\theta^6\left(27+16\sqrt{2\theta^2-1}\right)\right.\\
    &\left.+8\theta^4\left(9+32\sqrt{2\theta^2-1}\right)
    -8\theta^2\left(1+52\sqrt{2\theta^2-1}\right)\right]\\
    &-4\theta^6\log(\theta)\left[1-11\theta^2+27\theta^4-23\theta^6+10\theta^8\right.\\
    &\left.\left.+(1-2\theta^2)^2\,(1-2\theta^2+3\theta^4)\log(\theta)\right]\right) \; .
    \end{split}
\end{align}
%

\subsubsection*{RKL Loss}

For the expectation value of the gradient, we obtain for both directions
\begin{align}
    \nabla_\theta \loss_{\text{RKL}}^\text{fw}
    =\nabla_\theta \loss_{\text{RKL}}^\text{inv}
    =\frac{\theta^2-1}{\theta} \; .
\end{align}
While for the variance of the gradient, we obtain
\begin{align}
    \text{Var}\left(\nabla_\theta \loss_{\text{RKL}}^\text{fw}\right)
    &=\frac{21 - 54 \theta^2 + 37 \theta^4 + 4 \log(\theta)\,(3 - 5 \theta^2 + \log \theta)}{2 \theta^2}\\
    \text{Var}\left(\nabla_\theta \loss_{\text{RKL}}^\text{inv}\right)
    &=2\theta^2 \; .
\end{align}
In Fig.~\ref{fig:analytic_loss}, we illustrate the gradient variances for the forward and inverse training for the different loss functions.
The upper panels show the absolute gradient variance, while the lower panel shows the ratio between the forward and inverse directions. A ratio of $r>1$ means the gradient variance of the forward training is larger than the gradient variance of the inverse training. For the variance loss and KL divergence, we can observe that the forward training yields the more stable training. Only in parameter regions around the optimal value, \ie $\theta\approx \theta_\text{opt}=1$ the inverse training is more stable. In contrast, for the RKL divergence, the picture changes and the inverse loss gives less noisy gradients.

\section{Explicit channel mapping}
\label{app:channe_mapping_example}

\begin{figure}[tb!]
    \centering
    \includegraphics[height=3cm]{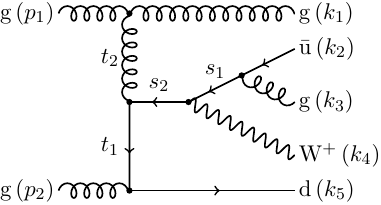}
    \hspace{10pt}
    \includegraphics[height=3cm]{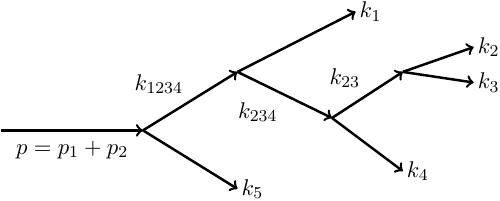}
    \caption{An example Feynman diagram contributing to the $\Pg \Pg \to \PW^{+} \Pubar \Pd \Pg \Pg$ process (left) and an illustration of the corresponding phase-space parametrization (right).}
    \label{fig:diagram_channel}
\end{figure}

As an example, we consider $\PW+4~\jets$ production
\begin{align}
\Pg \Pg \to \PW^{+} \Pubar \Pd \Pg \Pg\;.
\end{align}
In particular, we investigate the Feynman diagram of Fig.~\ref{fig:diagram_channel}, because it involves all types of phase-space blocks introduced in Sec.~\ref{sec:phase_space}. We define
\begin{align}
\begin{alignedat}{7}
    k_{23}&=k_2+k_3 \qquad & k_{234}&=k_2+k_3+k_4 \qquad & k_{1234}&=k_1+k_2+k_3+k_4 \\
    q_1 &= p_1 - k_1 \qquad & q_2 &= p_2 - k_5 \qquad & p &= p_1 + p_2\;.
\end{alignedat}
\end{align}
The infrared and collinear singularities are excluded by a lower cut on $k^2_{23}>k^2_{23,\text{min}}$. The phase-space integral
\begin{align}
\begin{split}
\int \d\Phi^{2\to 5}\big|_{\text{Fig.}\ref{fig:diagram_channel}}&=\int\limits_{k^2_{23,\text{min}}}^{\hat{s}}\d k^2_{23}
\int\limits_{k^2_{23}}^{\hat{s}}\d k^2_{234}
\int\limits_{k^2_{234}}^{\hat{s}}\d k^2_{1234}
\int \d \Phi^{2\to 2}(p_1,p_2;k^2_{1234}, k^2_{5})\\
&\times
\int \d \Phi^{2\to 2}(p_1,q_2;k^2_{1}, k^2_{234})
\int \d \Phi^{1\to 2}(k_{234};k^2_{23}, k^2_{4})
\int \d \Phi^{1\to 2}(k_{23};k^2_{2}, k^2_{3})\;.
\end{split}
\label{eq:ps_decomposition_example}
\end{align}
is decomposed into two $2\to 2$ scattering processes and two decays. The intermediate particles of this decomposition are the virtual particles with momenta $k_{23}$ and $k_{234}$, and an additional pseudo particle with momentum $k_{1234}$. 
First, we perform the luminosity sampling according to Eq.\eqref{eq:lumi_sampling} and obtain 
\begin{align}
\begin{split}
\hat{s}&=s^{z_0}_\text{lab}\hat{s}^{1-z_0}_\text{min} \qquad \mwith \qquad
\hat{s}_\text{min}=(M_\PW + k_{23,\text{min}})^2
\\
\xi&=\frac{1}{2}\log\frac{x_1}{x_2}\;,
\end{split}
\end{align}
needed to define $p_{1,2}$ and perform the boost into the proton-proton rest frame. Next, the invariant masses of the external particles of the scattering processes and particle decays have to be determined
\begin{align}
\begin{split}
    k^2_{23}&=\gbar^{\nu_1}_\text{prop}(z_1,0,k^2_{23,\text{min}},\hat{s})\\
    k^2_{234}&=\gbar^{\nu_2}_\text{prop}(z_2,0,k^2_{23},\hat{s})\\
    k^2_{1234}&=\gbar^{\nu=0}_\text{prop}(z_3,0,k^2_{234},\hat{s})\equiv z_3(\hat{s}-k^2_{234})+k^2_{234}\;.
\end{split}
\end{align}
While the time-like invariants $s_1=k^2_{23}$ and $s_2=k^2_{234}$ correspond to massless propagators in the diagram, $s_3=k^2_{1234}$ is the squared CM energy of the first $2\to 2$ scattering ($t_1$) and belongs to a pseudo particle. Consequently, $s_3$ is sampled flat. As a next step, the final-state momenta are calculated step-by-step:
\begin{enumerate}[label=(\roman*)]
    \item $p_1+p_2\to k_{1234} + k_5$:\\
    The initial-state gluons transform into the final-state $\Pd$-quark and a pseudo particle with momentum $k_{1234}$. The invariant mass of the $\Pd$-quark propagator is given by
    \begin{align}
        |q_1^2|=\gbar^{\nu_3}_\text{prop}(z_4,0,0,\hat{s}-k^2_{1234})
    \end{align}
    where the boundaries are taken from Eq.\eqref{eq:t_invariant_theta}.

    \item $p_1+q_2\to k_{1} + k_{234}$:\\
    The incoming particles are the initial-state gluon and the incoming virtual $\Pd$-quark. We further note that $q^2_2=(p_1-k_5)^2=t_1$. The invariant mass of the gluon propagator then reads
    \begin{align}
        |q_2^2|=\gbar^{\nu_4}_\text{prop}(z_5,0,0,(k^2_{1234}-k^2_{234})(k^2_{1234}-t_{1})/k^2_{1234})
    \end{align}
    where the boundaries are calculated again from Eq.\eqref{eq:t_invariant_theta}. This fixes the momenta $k_1$ and $k_{234}$ of the outgoing gluon and the virtual $\Pd$-quark, respectively.

    \item $k_{234}\to k_{23} + k_{4}$:\\
    The virtual $\Pdbar$-quark decays isotropically into the final-state W-boson ($k_4$) and the virtual $\Pubar$-quark ($k_{23}$).
    
    \item $k_{23}\to k_{2} + k_{3}$:\\
    Finally, the virtual $\Pubar$-quark decays isotropically into the final-state $\Pubar$-quark ($k_{2}$) and a gluon ($k_{3}$).
\end{enumerate}
The total phase-space density is then given by
\begin{align}
\begin{split}
g_\text{tot}&=
g_\text{lumi}\,(\tau,\tau_\text{min})
g^{\nu_1}_\text{prop}(k^2_{23},0,k^2_{23,\text{min}},\hat{s})
\,g^{\nu_2}_\text{prop}(k^2_{234},0,k^2_{23},\hat{s})
\,g^{\nu=0}_\text{prop}(k^2_{1234},0,k^2_{234},\hat{s})\\
&\times g_{2\to 2}(\hat{s},0,0,t_1,0,\nu_3,0,\hat{s}-k^2_{1234})\\
&\times g_{2\to 2}(k^2_{1234},0,t_1,t_2,0,\nu_4,0,(k^2_{1234}-k^2_{234})(k^2_{1234}-t_{1})/k^2_{1234})\\
&\times
g_\text{decay}(k^2_{234},k^2_{23},0)\,g_\text{decay}(k^2_{23},0,0)
\end{split}
\end{align}
which includes all propagators of this diagram.

\clearpage

\bibliography{tilman,refs}
\end{document}


\begin{center}{\Large \textbf{
Appendix to Differentiable \madnis}
}\end{center}

\begin{center}
Theo Heimel\textsuperscript{1},
Olivier Mattelaer\textsuperscript{2},
Tilman Plehn\textsuperscript{1},
and Ramon Winterhalder\textsuperscript{2}
\end{center}

\begin{center}
{\bf 1} Institut f\"ur Theoretische Physik, Universit\"at Heidelberg, Germany
\\
{\bf 2} CP3, Universit\'e catholique de Louvain, Louvain-la-Neuve, Belgium
\\
\end{center}

\begin{center}
\today
\end{center}


\section*{Abstract}
{\bf Probably not...}

\vspace{10pt}
\noindent\rule{\textwidth}{1pt}
\tableofcontents\thispagestyle{fancy}
\noindent\rule{\textwidth}{1pt}
\vspace{10pt}

\clearpage
\appendix

\section{Questions/ideas}

\begin{itemize}
    \item How many channels do we actually need?
    \begin{itemize}
        \item full multichannel setup
        \item RAMBO
        \item RAMBO + s-channel mappings (only resonant ones or all of them?)
        \item CHILI / CHILI basic
    \end{itemize}
    \item Build small trainable components
    \begin{itemize}
        \item Which correlations do you need to include to get the largest performance gains? Use interpretability of bilinear spline flows
        \item RAMBO: make parts of RAMBO trainable using its "autoregressive" structure.
    \end{itemize}
    \item Setups with low number of parameters:
    \begin{itemize}
        \item Are they faster to train or more sample efficient?
        \item See how high we can choose $R_@$ until the performance breaks down.
        \item Is it fast enough to be an alternative to VEGAS for small multiplicities?
        \item Better optimizers like L-BFGS feasible?
    \end{itemize}
\end{itemize}

\section{Bilinear spline flows}

\subsection{Architecture}

For typical \madnis trainings, we see that the flow sub-networks encode a relatively simple functional form. The idea behind bilinear spline flows is therefore to replace the sub-networks with a simple second-order polynomial. A $d_x$-dimensional transformation $x \leftrightarrow z$ with an $d_c$-dimensional condition $c$ can be written as
\begin{align}
    z = G(x; W \hat{c}) \qquad\mwith\qquad
    \hat{c} = (
        1, \; c_i \text{ for all } i,\; c_i c_j \text{ for all } i \leq j
    ) \; ,
\end{align}
where W is a trainable matrix. The number of trainable parameters $d_W$ for a spline transformation with $n_b$ bins is
\begin{align}
    d_W = d_x \: (3 n_b + 1) \: \left( 1 + d_c + \frac{1}{2} d_c (d_c+1) \right) \; .
\end{align}
This makes it possible to build very small, but still sufficiently expressive trainable transformations for low-dimensional $d_x$ and $d_c$. Because of the scaling $d_W = \mathcal{O}(d_c ^ 2)$, they become as expensive as neural networks in higher-dimensional cases, but are less expressive.

Another benefit of bilinear spline flows is their interpretability. We can directly read off from the the matrix $W$, how strongly the spline transformation is correlated with the conditional inputs.

\subsection{Results for WWW}

As an example, we look at the WWW-process from the last \madnis paper. We build a single-channel phase space integrator based on the most important channel identified by the trainable channel weights. This phase space mapping consists of four blocks:
\begin{itemize}
    \item the luminosity mapping, 2 random numbers
    \item a $t$-channel mapping that samples the invariant $t$ and the angle $\phi$, 2 random numbers
    \item an $s$-channel mapping, 1 random number
    \item a two-particle decay, samples the decay angles $\phi,\theta$, 2 random numbers
\end{itemize}
We compare four different setups:
\begin{enumerate}
    \item a full \madnis flow with the same hyperparameters as the last paper,
    \item a full \madnis flow, but with bilinear operations instead of sub-networks,
    \item a phasespace mapping enhanced with trainable components using bilinear flows,
    \item VEGAS as a baseline.
\end{enumerate}
For the third setup, we allow for correlations between the dimensions of blocks. In addition, the $t$-channel and $s$-channel mappings are conditioned on $\tau=\hat{s}/s$. All bilinear flows use RQ splines with three bins.

We run trainings using the four different setups and give the results in Tab.~\ref{tab:www_results}.

\begin{table}
    \centering
    \begin{tabular}{lrrr}
        \toprule
        Setup & \hspace{1em} Parameters & \hspace{1em} Unw. eff. [\%] & \hspace{1em} Rel. std. \\
        \midrule
        Full \madnis & 28683 & 50.0 & 0.153 \\
        \midrule
        \madnis, bilinear flows, 3 bins & 2550 & 36.0 & 0.257 \\
        Autoregressive Rambo, 3 bins & 610 & 20.2 & 0.563\\
        enhanced PS mapping, 6 bins & 513 & 15.2 & 0.934 \\
        enhanced PS mapping, 3 bins & 270 & 11.8 & 0.988 \\
        VEGAS & 889 & 8.6 & 1.208 \\
        \bottomrule
    \end{tabular}
    \caption{Training results for different setups applied to a single-channel WWW integral}
    \label{tab:www_results}
\end{table}

\section{Differentiable loss functions}

Consider an arbitrary $F$-divergence over a space $z$ between two normalized probability distributions $p_1(z)$ and $p_2(z)$,
\begin{align}
    D_F^z(p_1,p_2) = \int \d z p_2(z) \: F\left(\frac{p_1(z)}{p_2(z)}\right) \;.
\end{align}
For a \madnis-like training, we have a target function $f(x)$ and a normalizing flow that implements an invertible mapping
\begin{align}
x \quad 
\xleftrightarrow[\quad \leftarrow \gbar_\theta(y)\quad]{G_\theta(x)\rightarrow} 
\quad y
\label{eq:ps_mapping}
\end{align}
between a latent space $y \sim p(y)$ and the learned distribution
\begin{align}
    g_\theta(x) = p(G_\theta(x)) \left| \frac{\del G_\theta(x)}{\del x} \right| \; .
\label{eq:forward_prob}
\end{align}
For convenience, we also define
\begin{align}
\begin{split}
    \overline{g}_\theta(y) &= p(y) \left| \frac{\del \gbar_\theta(y)}{\del y} \right|^{-1} \\
    \text{such that}\quad \overline{g}_\theta(G_\theta(x)) &= g_\theta(x) \;.
\end{split}
\end{align}
Note that $g_\theta(x)$ is a normalized probability distribution in $x$-space (data space), but this is not the case for $\overline{g}_\theta(y)$ in $y$-space (latent space). There are two ways to define a loss function to train the flow. The first option is to define it using a divergence in data space,
\begin{align}
\begin{split}
    \mathcal{L}^\text{fw}_F &= D_F^x(f,g_\theta) \\
    &= \int \d x g_\theta(x) \: F\left(\frac{f(x)}{g_\theta(x)}\right) \\
    &= \left\langle \frac{g_\theta(x)}{q(x)} \: F\left(\frac{f(x)}{g_\theta(x)} \right) \right\rangle_{x \sim q(x)} \; .
\end{split}
\end{align}
In the last step, we have introduced an importance sampling distribution $q(x)$ that is used to evaluate the integral numerically. This can be the same as $g_\theta(x)$ (without taking its gradients into account), corresponding to online training, or a different distribution, enabling buffered training. Optimizing this loss function requires evaluating the flow in the forward direction according to Eq.~\eqref{eq:forward_prob}, so we refer to this training mode as forward training.

Alternatively, we can define the training in latent space. To this end, we define the remapped target distribution
\begin{align}
    \hat{f}(y) = f(\gbar_\theta(y)) \left| \frac{\del \gbar_\theta(y)}{\del y} \right| \; ,
\end{align}
which is a normalized probability in latent space according to the change of variables formula. Now our training objective is to minimize the divergence between this remapped distribution and the latent space distribution,
\begin{align}
\begin{split}
    \mathcal{L}^\text{inv}_F &= D_F^y(\hat{f},p) \\
    &= \int \d y \,p(y) \: F\left(\frac{\hat{f}(y)}{p(y)}\right) \\
    &= \int \d y \,p(y) \: F\left(\frac{f(\gbar_\theta(y))}{p(y)} \left| \frac{\del \gbar_\theta(y)}{\del y} \right| \right) \\
    &= \int \d y \,p(y) \: F\left(\frac{f(\gbar_\theta(y))}{\overline{g}_\theta(y)} \right) \\
    &= \left\langle F\left(\frac{f(\gbar_\theta(y))}{\overline{g}_\theta(y)} \right) \right\rangle_{y \sim p(y)} \; .
\end{split}
\end{align}
We refer to this training mode as inverse training because we evaluate the flow in the inverse direction.

We can show that these two ways of defining the loss function yield the same result,
\begin{align}
\begin{split}
    \mathcal{L}^\text{inv}_F &= \int \d y \,p(y) \: F\left(\frac{f(\gbar_\theta(y))}{\overline{g}_\theta(y)} \right) \\
    &= \int \d x \,p(G_\theta(x)) \left| \frac{\del G_\theta(x)}{\del x} \right| \: F\left(\frac{f(\gbar_\theta(G_\theta(x)))}{\overline{g}_\theta(G_\theta(x))} \right) \\
    &= \int \d x \,g_\theta(x) \: F\left(\frac{f(x)}{g_\theta(x)} \right)
    = \mathcal{L}^\text{fw}_F \; .
\end{split}
\end{align}
Consequently, we can also show that the gradients of these loss functions have the same expectation value,
\begin{align}
\begin{split}
    \left\langle \nabla_\theta F\left(\frac{f(\gbar_\theta(y))}{\overline{g}_\theta(y)} \right) \right\rangle_{y \sim p(y)}
    &= \nabla_\theta \left\langle F\left(\frac{f(\gbar_\theta(y))}{\overline{g}_\theta(y)} \right) \right\rangle_{y \sim p(y)} \\
    &= \nabla_\theta \mathcal{L}^\text{inv}_F \\
    &= \nabla_\theta \mathcal{L}^\text{fw}_F \\
    &= \nabla_\theta \left\langle \frac{g_\theta(x)}{q(x)} \: F\left(\frac{f(x)}{g_\theta(x)} \right) \right\rangle_{x \sim q(x)} \\
    &= \left\langle \nabla_\theta \frac{g_\theta(x)}{q(x)} \: F\left(\frac{f(x)}{g_\theta(x)} \right) \right\rangle_{x \sim q(x)} \; .
\end{split}
\end{align}
However, this is not true for the variances of the gradients,
\begin{align}
    \text{Var}_{x\sim q(x)} \left( \nabla_\theta \frac{g_\theta(x)}{q(x)} \: F\left(\frac{f(x)}{g_\theta(x)} \right) \right)
    &= \left\langle \left( \nabla_\theta \frac{g_\theta(x)}{q(x)} \: F\left(\frac{f(x)}{g_\theta(x)} \right) - \nabla_\theta\mathcal{L}^\text{fw}_F \right)^2 \right\rangle_{x\sim q(x)}\notag\\
    &=\int \d x\,q(x) \left( \nabla_\theta \frac{g_\theta(x)}{q(x)} \: F\left(\frac{f(x)}{g_\theta(x)} \right) - \nabla_\theta\mathcal{L}^\text{fw}_F \right)^2\;,\\[1em]
    \text{Var}_{y\sim p(y)} \left( \nabla_\theta F\left(\frac{f(\gbar_\theta(y))}{\overline{g}_\theta(y)} \right) \right)
    &= \left\langle \left( \nabla_\theta F\left(\frac{f(\gbar_\theta(y))}{\overline{g}_\theta(y)} \right) - \nabla_\theta\mathcal{L}^\text{inv}_F \right)^2 \right\rangle_{y\sim p(y)}\notag\\
    &= \int \d y\, p(y) \left( \nabla_\theta F\left(\frac{f(\gbar_\theta(y))}{\overline{g}_\theta(y)} \right) - \nabla_\theta\mathcal{L}^\text{inv}_F \right)^2 \;.
\end{align}
In general, these will not be equal. Consequently, depending on the problem and choice of $F$-divergence, one of the two training modes can have noisier gradients, leading to a slower convergence of the training on finite batch sizes.

Finally, we can look at some concrete examples of loss functions following this construction. We assume a normalized target distribution $f$, which can be ensured during training using batchwise normalization of the integrand values.
\begin{itemize}
    \item variance, $F(t) = (t-1)^2$
    \begin{align}
        \mathcal{L}^\text{fw}_\text{var} &= \left\langle \frac{g_\theta(x)}{q(x)} \: \left(\frac{f(x)}{g_\theta(x)} - 1\right)^2 \right\rangle_{x \sim q(x)} \\
        \mathcal{L}^\text{inv}_\text{var} &= \left\langle \left(\frac{f(\gbar_\theta(y))}{\overline{g}_\theta(y)} - 1 \right)^2 \right\rangle_{y \sim p(y)}
    \end{align}
    \begin{figure}[H]
    \centering
    \includegraphics[width=0.49\textwidth]{figs_old/var_ratio_variance.pdf}
    \caption{Ratio of gradient variances for the Variance loss.}
    \label{fig:gradvar_variance}
    \end{figure}
    \item KL-divergence, $F(t) = t \log t$
    \begin{align}
        \mathcal{L}^\text{fw}_\text{KL} &= \left\langle \frac{f(x)}{q(x)} \: \log \frac{f(x)}{g_\theta(x)} \right\rangle_{x \sim q(x)} \\
        \mathcal{L}^\text{inv}_\text{KL} &= \left\langle \frac{f(\gbar_\theta(y))}{\overline{g}_\theta(y)} \log \frac{f(\gbar_\theta(y))}{\overline{g}_\theta(y)} \right\rangle_{y \sim p(y)}
    \end{align}
    \begin{figure}[H]
    \centering
    \includegraphics[width=0.49\textwidth]{figs_old/var_ratio_kl.pdf}
    \caption{Ratio of gradient variances for the KL loss.}
    \label{fig:gradvar_KL}
    \end{figure}
    \item reverse KL-divergence, $F(t) = -\log t$
    \begin{align}
        \mathcal{L}^\text{fw}_\text{RKL} &= \left\langle \frac{g_\theta(x)}{q(x)} \: \log\frac{g_\theta(x)}{f(x)} \right\rangle_{x \sim q(x)} \\
        \mathcal{L}^\text{inv}_\text{RKL} &= \left\langle \log\frac{\overline{g}_\theta(y)}{f(\gbar_\theta(y))} \right\rangle_{y \sim p(y)}
    \end{align}
    \begin{figure}[H]
    \centering
    \includegraphics[width=0.49\textwidth]{figs_old/var_ratio_rkl.pdf}
    \caption{Ratio of gradient variances for the RKL loss.}
    \label{fig:gradvar_RKL}
    \end{figure}
\end{itemize}


%

\subsection{Score matching}
When using differentiable integrands we can also implement a score matching loss
\begin{align}
    \mathcal{L}_\text{score}=\left\langle|\overline{\partial_x\log f_{i\xi}(x)} - \partial_x g_{i\theta}(x)|^2\right\rangle_{x\sim q_i}
\end{align}

\subsection{Analytic example}
In order to exemplify the properties of the forward and inverse loss, we consider a simple 1-dimensional target function $f$ as
\begin{align}
    f(x)=\frac{1}{\sqrt{2\pi}}\exp\left(-\frac{x^2}{2}\right)\;.
\end{align}
Our prior distribution is given as a standard normal distribution as
\begin{align}
    p_0(z)=\frac{1}{\sqrt{2\pi}}\exp\left(-\frac{z^2}{2}\right)\;.
\end{align}
We now assume a simple trainable mapping (1d flow if you want), given by
\begin{align}
    x\equiv \overline{G}_a(z)=az\quad\longleftrightarrow \quad z\equiv G_a(x)=\frac{x}{a}\;.
\end{align}
inducing the jacobian determinants
\begin{align}
    \left\vert\frac{\partial G_a(x)}{\partial x}\right\vert=\frac{1}{a} \quad \longleftrightarrow \quad
    \left\vert\frac{\partial \overline{G}_a(z)}{\partial z}\right\vert=a\;.
\end{align}
Combining this with the prior sample density, this yields the overall importance sampling density
\begin{align}
    g_a(x)&=p_0(G_a(x))\left\vert\frac{\partial G(x)}{\partial x}\right\vert=\frac{1}{\sqrt{2\pi a^2}}\exp\left(-\frac{x^2}{2a^2}\right)\\[0.2cm]
    \overline{g}_a(z)&=p_0(z)\left\vert\frac{\partial \overline{G}_a(z)}{\partial z}\right\vert^{-1}=\frac{1}{\sqrt{2\pi a^2}}\exp\left(-\frac{z^2}{2}\right)\qquad \mwith\quad \overline{g}_a(G_a(x)) = g_\theta(x) \;.
\end{align}
In this case the integral of $f$ can be calculated as
\begin{align}
    I_f&=\int\d x\;f(x)\notag\\
    &=\int\d x\;g_a(x)\,\frac{f(x)}{g_a(x)}\notag\\
    &=\left\langle\frac{g_a(x)}{q(x)}\,\frac{f(x)}{g_a(x)}\right\rangle_{x\sim q(x)}=1\;.
\end{align}
\subsubsection{Forward Variance loss}
While the integral does not change, the variance of the integrand is given by
\begin{align}
\begin{split}
    \mathbb{V}^{\rightarrow}_f&=\left\langle\frac{g_a(x)}{q(x)}\,\left(\frac{f(x)} {g_a(x)}-I_f\right)^2\right\rangle_{x\sim q(x)} \\
    &=\int\d x\,g_a(x)\,\left(\frac{f(x)}{g_a(x)}-I_f\right)^2  \\
    &=\frac{a^2}{\sqrt{2a^2-1}}-1
\end{split}
\end{align}
which is exactly zero for the expected value of $a=1$. Next we can calculate the expectation value of the gradient of the integrand with respect to $a$ as this quantity is used during optimization. 
This yields
\begin{align}
\begin{split}
\left\langle\nabla_a\frac{g_a(x)}{q(x)}\,\left(\frac{f(x)}{g_a(x)}-I_f\right)^2\right\rangle_{x\sim q(x)}
&=\nabla_a\,\left\langle\frac{g_a(x)}{q(x)}\,\left(\frac{f(x)}{g_a(x)}-I_f\right)^2\right\rangle_{x\sim q(x)}\\
&=\nabla_a \mathbb{V}^{\rightarrow}_f\\
&=\frac{2a\,(a^2-1)}{(2 a^2-1)^{3/2}}
\end{split}
\end{align}
On the other hand, the variance of the gradient is given by
\begin{align}
\begin{split}
\text{Var}^{\rightarrow}_{a}\left(\nabla_a\frac{g_a(x)}{q(x)}\,\left(\frac{f(x)}{g_a(x)}-I_f\right)^2\right)
&=\left\langle\left(\nabla_a\frac{g_a(x)}{q(x)}\,\left(\frac{f(x)}{g_a(x)}-I_f\right)^2-\nabla_a \mathbb{V}^{\rightarrow}_f\right)^2\right\rangle_{x\sim q(x)}\\
&=\int \d x \,q(x)\left(\nabla_a\frac{g_a(x)}{q(x)}\,\left(\frac{f(x)}{g_a(x)}-I_f\right)^2-\nabla_a \mathbb{V}^{\rightarrow}_f\right)^2\\
&=-\frac{1}{4}\left(2-\frac{8}{a^2}-\frac{2}{(2a^2-1)^2}+\frac{2}{(2a^2-1)^3}+\frac{24}{(2a^2-1)^{5/2}}\right.\\
&-\frac{16}{(2a^2-1)^{3/2}}-\frac{2}{2a^2-1}+\frac{8}{(2a^2-1)^{1/2}}\\
&-\frac{9}{(4a^2-3)^{3/2}}+\frac{3}{(4a^2-3)^{3/2}}\\
&\left.-\frac{1}{(4a^2-3)^{1/2}}-\sqrt{4a^2-3} \right)
\end{split}
\end{align}

\subsubsection{Inverse Variance loss}
In the inverse direction, we obtain the variance as
\begin{align}
\begin{split}
    \mathbb{V}^{\leftarrow}_f&=\left\langle\left(\frac{f(\overline{G}_a(z))}{\overline{g}_a(z)}-I_f\right)^2\right\rangle_{z\sim p_0(z)} \\
    &=\int\d z\,p_0(z)\,\left(\frac{f(\overline{G}_a(z))}{\overline{g}_a(z)}-I_f\right)^2  \\
    &=\frac{a^2}{\sqrt{2a^2-1}}-1
\end{split}
\end{align}
which yields the same variance, as expected.
Similarly, the expectation value of the gradient yields
\begin{align}
\begin{split}
\left\langle\nabla_a\left(\frac{f(\overline{G}_a(z))}{\overline{g}_a(z)}-I_f\right)^2\right\rangle_{z\sim p_0(z)}
&=\nabla_a\,\left\langle\left(\frac{f(\overline{G}_a(z))}{\overline{g}_a(z)}-I_f\right)^2\right\rangle_{z\sim p_0(z)}\\
&=\nabla_a \mathbb{V}^{\leftarrow}_f =\nabla_a \mathbb{V}^{\rightarrow}_f\\
&=\frac{2a\,(a^2-1)}{(2 a^2-1)^{3/2}}
\end{split}
\end{align}
However, for the variance we obtain
\begin{align}
\begin{split}
\text{Var}^{\rightarrow}_{a}\left(\nabla_a\left(\frac{f(\overline{G}_a(z))}{\overline{g}_a(z)}-I_f\right)^2\right)
&=\left\langle\left(\nabla_a\left(\frac{f(\overline{G}_a(z))}{\overline{g}_a(z)}-I_f\right)^2-\nabla_a \mathbb{V}^{\leftarrow}_f\right)^2\right\rangle_{z\sim p_0(z)}\\
&=\int \d z \,p_0(z)\left(\nabla_a\left(\frac{f(\overline{G}_a(z))}{\overline{g}_a(z)}-I_f\right)^2-\nabla_a \mathbb{V}^{\leftarrow}_f\right)^2\\
&=-\frac{1}{16}\left(8-\frac{8}{(2a^2-1)^2}+\frac{8}{(2a^2-1)^3}-\frac{48}{(2a^2-1)^{5/2}}\right.\\
&-\frac{32}{(2a^2-1)^{3/2}}-\frac{8}{2a^2-1}-\frac{48}{(2a^2-1)^{1/2}}\\
&-\frac{81}{(4a^2-3)^{3/2}}-\frac{9}{(4a^2-3)^{3/2}}-\frac{27}{(4a^2-3)^{1/2}}\\
&\left.-11\sqrt{4a^2-3}+\frac{256a\,(3 a^4- 4 a^2+2)}{(3a^2-2)^{5/2}} \right)
\end{split}
\end{align}

\subsubsection{KL Loss}

For the expectation value of the gradient we obtain for both directions
\begin{align}
    \mathbb{E}^{\to}\left(\nabla_a \mathcal{L}_{\text{KL}}\right)
    =\mathbb{E}^{\leftarrow}\left(\nabla_a \mathcal{L}_{\text{KL}}\right)
    =\frac{a^2-1}{a^3}
\end{align}
While vor the variance we obtain
\begin{align}
    \text{Var}_a^{\to}\left(\nabla_a \mathcal{L}_{\text{KL}}\right)
    &=\frac{21 - 54 a^2 + 37 a^4 + 4 \log(a)\,(3 - 5 a^2 + \log a)}{2 a^2}\\
    \text{Var}_a^{\leftarrow}\left(\nabla_a \mathcal{L}_{\text{KL}}\right)
    &=\frac{-1}{4a^6(2a^2-1)^{9/2}}\left(
    \sqrt{2a^2-1}\,\left(4-40a^2+164a^4\right)\right.\notag\\
    &-a^6\left[3+49a^8+352\sqrt{2a^2-1}-4a^6\left(27+16\sqrt{2a^2-1}\right)\right.\notag\\
    &\left.+8a^4\left(9+32\sqrt{2a^2-1}\right)
    -8a^2\left(1+52\sqrt{2a^2-1}\right)\right]\notag\\
    &-4a^6\log(a)\left[1-11a^2+27a^4-23a^6+10a^8\right.\notag\\
    &\left.\left.+(1-2a^2)^2\,(1-2a^2+3a^4)\log(a)\right]\right)
\end{align}

\subsubsection{RKL Loss}

For the expectation value of the gradient we obtain for both directions
\begin{align}
    \mathbb{E}^{\to}\left(\nabla_a \mathcal{L}_{\text{RKL}}\right)
    =\mathbb{E}^{\leftarrow}\left(\nabla_a \mathcal{L}_{\text{RKL}}\right)
    =\frac{a^2-1}{a}
\end{align}
While vor the variance we obtain
\begin{align}
    \text{Var}_a^{\to}\left(\nabla_a \mathcal{L}_{\text{RKL}}\right)
    &=-\frac{1}{a^6}+\frac{2}{a^4}-\frac{1}{a^2} + \frac{4 a^4 - 8 a^2 + 6}{(2a^2-1)^{5/2}}\\
    \text{Var}_a^{\leftarrow}\left(\nabla_a \mathcal{L}_{\text{RKL}}\right)
    &=2a^2
\end{align}

\subsection{Generalized KL}
We can also define a KL divergence which is generalized and does not rely on normalized distributions. This can be defined as \cite{miller2023simulationbased}
\begin{align}
    \mathcal{L}_\text{GKL}(f_{i\xi}||g_{i\theta})&:=\int \d x\;\phi\!\left(\frac{g_{i\theta}(x)}{f_{i\xi}(x)}\right)\,f_{i\xi}(x)\,,
    \qquad \mwith \quad
    \phi(r)=-\log r + r -1\notag\\
    &=\int \d x\;f_{i\xi}(x)\,\log\frac{f_{i\xi}(x)}{g_{i\theta}(x)} + \int \d x\;g_{i\theta}(x) - \int \d x\;f_{i\xi}(x)\notag\\
    &=\left\langle \frac{f_{i\xi}(x)}{q(x)}\,\log\frac{f_{i\xi}(x)}{g_{i\theta}(x)}\right\rangle_{x\sim q} - I_{i\xi} +\text{const}
\end{align}

\subsection{Remarks on MadJax}

The loss function used in MadJax to train flows is given by
\begin{align}
\begin{split}
    \mathcal{L} &= \mathcal{L}_{\text{NLL}}
    + \lambda_k \mathcal{L}_{\text{RKL}}
    + \lambda_f \mathcal{L}_{\text{FM}}\\
    \mwith\quad \mathcal{L}_{\text{NLL}} &= - \left\langle \log g_\theta(x) \right\rangle_{x \sim p(x)} \;, \\
    \mathcal{L}_{\text{RKL}} &= \left\langle \log \frac{g_\theta(x)}{p(x)} \right\rangle_{x \sim g_\theta(x)} \;, \\
    \mathcal{L}_{\text{FM}} &= \left\langle |\partial_x\log p(x) - \partial_x g_\theta(x)|^2 \right\rangle_{x \sim p(x)} \;.
\end{split}
\end{align}
The expectation values over $x\sim p(x)$ were evaluated for a fixed set of 8000 events, with a batch size of 100. The networks were trained for 200 epochs. It is not exactly stated how many samples are used for the reverse KL loss. Another 100 samples per batch would be a reasonable choice. The authors observe an improvement from introducing the RKL loss term and another small improvement from the force matching loss (called score matching above).

It is not clear whether the improvements from introducing the RKL loss term are actually coming from the use of gradient information. It seems more likely that the improvement is just coming from the dramatic increase in training statistics, as 8000 events are a very small training dataset for a process with a six-particle final state. Over the whole training, the RKL loss term contributes $200 \times 8000 = 1.6\text{M}$ additional (weighted) samples to the training.

\subsection{Remarks on buffered training}

In a realistic \madnis training, especially for complicated processes, we want to perform most network weight updates using buffered samples. Therefore the question is how the use of gradient information could be generalized to training on buffered samples without having to evaluate the integrand again. In the simplest case without trainable channel weights, we could store the gradient $\partial_x \left[\alpha_i(x) f(x)\right]$ (a $d$-dimensional vector). By first evaluating the flow on a sample $x$ in the forward direction without gradients to obtain the latent space $z$, we could then perform an inverse loss weight update using the stored gradient. Furthermore, the score matching loss could also be computed using the stored gradient.

The situation becomes more complicated once we introduce trainable channel weights. The channel weights depend on the prior channel weights, which are computed during the integrand evaluation. To compute their gradients without having to compute the prior channel weights again, we would need to store the gradients $\partial_x \alpha^\text{MG}_i(x)$ for all $i$. For every sample, this would be a matrix of size $d \times n_c$ which would quickly become prohibitive for processes with many channels. Note that just storing $\alpha^\text{MG}_i(x)$ alone for all $i$ is currently the most memory-intensive part of buffered training. Storing the full Jacobian would use another $d$ times more memory than that.

Hence, buffered training with gradient information and trainable channel weights is not feasible and we would have to choose between one of those features.

\section{Toy examples}

\subsection{Broad multi-dimensional camel, single channel}

In this example we consider a $d$-dimensional camel with $m$ modes, as given by
\begin{align}
\label{eq:camel_function}
    f_{m}^{(n)}(\vec{x})
        = \sum_{i=1}^m \frac{\omega_i}{\left(\sigma_i\sqrt{2\pi}\right)^{d}}\,\exp\left[-\frac{1}{2}\frac{ \vert\vec{x}-\vec{\mu}_i\vert^2}{\sigma^2_i}\right], \quad \mathrm{with} \quad \sum_i \omega_i = 1.
\end{align}
which integrated on the unit-hypercube $[0,1]^d$ has the integral
\begin{align}
\label{eq:camel_integral}
    I_{m}^{(d)} = \int\limits_{0}^{1}\mathrm{d}^d x\ f_{m}^{(d)}(\vec{x})=\sum_{i=1}^m\ \omega_i\ \prod_{j=1}^d\left(\frac{1}{2}\left[\mathrm{erf}\left(\frac{1-\mu^j_i}{\sqrt{2}\sigma^j_i}\right)+\mathrm{erf}\left(\frac{\mu^j_i}{\sqrt{2}\sigma^j_i}\right)\right]\right).
\end{align}
We choose $m=2$, $d=10$, $\omega_1 = \omega_2 = 0.5$, $\mu_1^i = 1/3$, $\mu_2^i = 2/3$, $\sigma_1^i = \sigma_2^i = 0.2$ for all $i=1,\ldots,d$. We normalize the function from Eq.~\eqref{eq:camel_function} using Eq.~\eqref{eq:camel_integral} such that the integral of our target function is 1.

\subsubsection*{Variance, KL, reverse KL}

We start with a long training with constant learning rate to determine after how many iterations the training is close to convergence, see Fig.~\ref{fig:loss_long}. The training is mostly converged after 100 epochs with 100 batches of 1024 samples. After that, the training with an inverse loss function becomes instable. In the following, we will look at 100-epoch trainings with constant learning rate.

\begin{figure}
    \centering
    \includegraphics[width=0.49\textwidth]{figs_old/loss_long.pdf}
    \caption{Losses for a 200 epoch training with constant learning rate. The networks were trained with a variance loss and forward and inverse flow evaluation. We show the median of 10 independent trainings and the range between the second-to-lowest and second-to-largest loss as an error estimate.}
    \label{fig:loss_long}
\end{figure}

Next, we look at different types of loss functions: variance, KL divergence and reverse KL divergence. For each option, we run a forward and inverse training. We show the training loss in Fig.~\ref{fig:loss_long} and the standard deviation and unweighting efficiency after the training in Fig.~\ref{fig:loss_types}. We find the best results with the variance loss. For the variance and KL divergence loss, the inverse training leads to a worse performance and more instability. Only for the reverse KL divergence, it leads to an improvement in both performance and stability.

\begin{figure}
    \centering
    \includegraphics[width=0.32\textwidth]{figs_old/loss_variance.pdf}
    \includegraphics[width=0.32\textwidth]{figs_old/loss_kl.pdf}
    \includegraphics[width=0.32\textwidth]{figs_old/loss_rkl.pdf}
    \caption{Losses for 100 epoch trainings with constant learning rate. The networks were trained with variance, KL and reverse KL losses and forward and inverse flow evaluation. We show the median of 10 independent trainings and the range between the second-to-lowest and second-to-largest loss as an error estimate.}
    \label{fig:loss_types}
\end{figure}

\begin{figure}
    \centering
    \includegraphics[width=0.49\textwidth]{figs_old/stddev_losses.pdf}
    \includegraphics[width=0.49\textwidth]{figs_old/uweff_losses.pdf}
    \caption{Final standard deviations and unweighting efficiencies for 100 epoch trainings with constant learning rate. The networks were trained with variance, KL and reverse KL losses and forward and inverse flow evaluation. We show the median of 10 independent trainings and the standard deviation as an error estimate.}
    \label{fig:metrics_losses}
\end{figure}

\subsubsection*{Score matching}

Next, we add a score matching term to the the variance loss. We show the results in Figs.~\ref{fig:loss_score} and~\ref{fig:metrics_score}. There is no clear improvement from the score matching and it even makes the results worse compared to the forward variance training without score matching.



We repeat these experiments with KL and RKL losses and again see no improvement from the score matching, see Figs.~\ref{fig:loss_score_kl} and~\ref{fig:metrics_score_kl}.

\begin{figure}
    \centering
    \includegraphics[width=0.49\textwidth]{figs_old/loss_forward_var_score.pdf}
    \includegraphics[width=0.49\textwidth]{figs_old/loss_inverse_var_score.pdf}\\
    \includegraphics[width=0.49\textwidth]{figs_old/loss_forward_kl_score.pdf}
    \includegraphics[width=0.49\textwidth]{figs_old/loss_inverse_kl_score.pdf}\\
    \includegraphics[width=0.49\textwidth]{figs_old/loss_forward_rkl_score.pdf}
    \includegraphics[width=0.49\textwidth]{figs_old/loss_inverse_rkl_score.pdf}
    \caption{Losses for 100 epoch trainings with constant learning rate. The networks were trained with a KL/RKL loss combined with score matching losses with different prefactors, and forward and inverse flow evaluation. We show the median of 10 independent trainings and the range between the second-to-lowest and second-to-largest loss as an error estimate.}
    \label{fig:loss_score}
\end{figure}

\begin{figure}
    \centering
    \includegraphics[width=0.49\textwidth]{figs_old/stddev_score.pdf}
    \includegraphics[width=0.49\textwidth]{figs_old/uweff_score.pdf}
    \caption{Final standard deviations and unweighting efficiencies for 100 epoch trainings with constant learning rate. The networks were trained a KL/RKL loss combined with score matching losses with different prefactors, and forward and inverse flow evaluation. We show the median of 10 independent trainings and the standard deviation as an error estimate.}
    \label{fig:metrics_score}
\end{figure}

\subsection{Narrow multi-dimensional camel, single channel}

To make our toy problem more difficult, we increase the distance of the camel peaks and make them more narrow: $\mu_1^i = 1/4$, $\mu_2^i = 3/4$, $\sigma_1^i = \sigma_2^i = 0.12$. We repeat our experiments from above and observe the following behavior:
\begin{itemize}
    \item Forward variance: Training converges and integration yields the corrects results.
    \item Inverse variance: With the same settings as before, the training does not converge. The resulting integral is zero because the network pushes all samples into a corner of the hypercube. This is probably caused by noisy gradients, as we can get the training to converge by increasing the batch size from 1024 to 10240.
    \item Forward KL divergence: The training converges, but there is mode collapse.
    \item Inverse KL divergence: Like for the inverse variance, the training does not converge. It only starts to converge after increasing the batch size by a factor 100.
    \item Forward RKL divergence: The training converges, but there is mode collapse.
    \item Inverse RKL divergence: The training converges, but there is mode collapse.
\end{itemize}
This confirms our previous observations that the training is made less stable by using inverse training combined with the variance or KL loss.

\subsection{Narrow multi-dimensional camel, two channels}

As we saw mode collapse when for the single channel training with the narrow camel distribution, we now move to a two-channel setup. We construct two channels with the mapping
\begin{align}
    y = \gbar_{1,2}(z) = \text{sigmoid}(\text{logit}(z) \pm 1) \; ,
\end{align}
applied to each component of the vector $z$. We combine this with a \sherpa prior,
\begin{align}
    \alpha_{1,2}^\text{prior}(x) = \frac{g_{1,2}(x)}{g_1(x) + g_2(x)} \; .
\end{align}
Each of the two mappings prefers one half of phase-space, so it should prevent mode collapse. At the same time, the mappings are still relatively wide, so the network has a non-trivial learning task.

\subsection{To-do list}

\begin{itemize}
    \item Play with LR, LR scheduling, etc. to see if there are settings that improve the inverse training results.
    \item Explore larger range of score matching coefficients.
    \item Build a combined loss with variance and RKL, more similar to the one used in MadJax.
    \item Check if there are differences between normal and generalized KL in a multichannel setting.
    \item Combine score matching with KL and RKL loss.
    \item Study mode collapse for camel with more separation.
    \item Test $\alpha$-divergence with some $\alpha>2$. This should, in theory, penalize large weights even stronger than the variance and result in higher unweighting efficiencies.
\end{itemize}

\section{Physics example}

\subsection{Already implemented}

\begin{itemize}
    \item Finish diff. matrix elements $\to$ done $\checkmark$ and pushed to \texttt{MadTorch} repo
    \item Read params from param card $\to$ can be read from arbitrary dictionary $\checkmark$
    \item Differentiable Rambo and RamboOnDiet $\to$ $\checkmark$ and pushed to \texttt{MadSprace} repo
    \item Differentiable PS-Mappping for channels $\to$ $\checkmark$ and pushed to \texttt{MadSprace} repo
\end{itemize}

\subsection{Single channel with RAMBO}

\section{Potential Beyond-MadNIS applications}
\begin{itemize}
    \item Calculation of profiled likelihoods (in MEM: maximize likelihood as a function of $x_\text{hard}$ instead of integrating it out)
    \item improved surrogate/amplitude learning (score matching)
    \item MadMiner (score matching)
    \item Annealed importance sampling (in MadNIS: run a few Langevin MC steps on the generated samples to improve the unweighting efficiency)
    \item ....
\end{itemize}

\section{Generalized loss functions}

The variance, KL and reverse KL losses introduced above can be further generalized by describing them as examples from a continuous spectrum of loss functions. We define the $\alpha$-variance and $\alpha$-divergence as
\begin{align}
    F^\text{var}_\alpha &= |t - 1|^\alpha \\
    F^\text{div}_\alpha &= \begin{cases}
    \frac {t^\alpha-\alpha t-\left(1-\alpha \right)}{\alpha \left(\alpha -1\right)} & \text{if } \alpha \neq 0,\;\alpha \neq 1,\\
    t\ln t-t+1,&\text{if } \alpha = 1,\\
    -\ln t+t-1,&\text{if } \alpha = 0 \; .
    \end{cases}
\end{align}
For the latter, the cases $\alpha=0$ and $\alpha=1$ correspond to the KL and RKL divergence, respectively.

\subsection{Results}

Results for 200-epoch trainings with a generalized variance loss are shown in Fig.~\ref{fig:alphavar}. The best results, both in the variance and unweighting efficiency, are achieved with $\alpha=1.5$. In the early stages of the training, losses with a lower $\alpha$ converge faster. In later stages of the training, they are flatter. Overall, lower $\alpha$ lead to a higher stability of the loss.

\begin{figure}[tb]
    \centering
    \includegraphics[width=0.49\textwidth]{figs_old/alphavar_stddev.pdf}
    \includegraphics[width=0.49\textwidth]{figs_old/alphavar_uweff.pdf}
    \includegraphics[width=0.49\textwidth]{figs_old/alphavar_loss.pdf}
    \caption{Final standard deviation, final unweighting efficiency and loss functions for networks trained with an $\alpha$-variance loss and different values of $\alpha$.}
    \label{fig:alphavar}
\end{figure}

\subsection{Stratified $\alpha$-variance}

So far, we have only looked at the single-channel case. To apply the generalized loss functions in a multi-channel setup with trainable channel weights, we have to derive a loss function similar to the stratified \madnis loss.

Given channel weights $\alpha_i(x)$, mappings $g_i(x)$ and a target function $f(x)$, we can write the estimate for the integral as
\begin{align}
\begin{split}
    \hat{I} &= \sum_{i=1}^{n_c} \frac{1}{N_i} \sum_{j=1}^{N_i}
    \frac{\alpha_i(x_{ij}) f(x_{ij})}{g_i(x_{ij})} \\
    &\equiv \sum_{i=1}^{n_c} \frac{1}{N_i} \sum_{j=1}^{N_i} w_{ij}
    \quad\mwith\quad x_{ij} \sim g_i(x_{ij}) \; .
\end{split}
\end{align}
Let $I$ be the true integral and $I_i$ be the channel-wise true integrals for given channel weights $\alpha_i(x)$. We can then write the $\alpha$-variance of the estimated integral as
\begin{align}
\begin{split}
    \text{Var}_\alpha(\hat{I})
    &= \left\langle \left| \hat{I} - \left\langle \hat{I} \right\rangle \right|^\alpha \right\rangle
    = \left\langle \left| \hat{I} - I \right|^\alpha \right\rangle \\
    &= \left\langle \left| \sum_{i=1}^{n_c} \frac{1}{N_i} \sum_{j=1}^{N_i} w_{ij} - I \right|^\alpha \right\rangle \\
    &= \left\langle \left| \sum_{i=1}^{n_c} \left[ \frac{1}{N_i} \sum_{j=1}^{N_i} w_{ij} - I_i \right] \right|^\alpha \right\rangle \\
    &= \left\langle \left| \sum_{i=1}^{n_c} \sum_{j=1}^{N_i} \frac{1}{N_i} (w_{ij} - I_i) \right|^\alpha \right\rangle
\end{split}
\end{align}
The $w_{ij}$ are independent random variables. In the case of the regular variance ($\alpha = 2$), we could express the variance of the sum as the sum over the variances. In the more general case, this is not easily possible. However, for $1 \leq \alpha \leq 2$, we can use the inequality derived in Ref.~\cite{10.1214/aoms/1177700291} to derive an upper bound,
\begin{align}
\label{eq:alpha_var_bound}
\begin{split}
    \text{Var}_\alpha(\hat{I})
    &\leq 2 \sum_{i=1}^{n_c} \sum_{j=1}^{N_i} \left\langle \left| \frac{1}{N_i} (w_{ij} - I_i) \right|^\alpha \right\rangle \\
    &= 2 \sum_{i=1}^{n_c} \frac{1}{N_i^\alpha} N_i \left\langle \left| (w_i - I_i) \right|^\alpha \right\rangle \\
    &= 2 \sum_{i=1}^{n_c} N_i^{1-\alpha} \; \text{Var}_\alpha(\hat{I}_i) \; .
\end{split}
\end{align}
We can now derive the $N_i$ that minimize this upper bound for the $\alpha$-variance by writing it as an optimization problem with a boundary condition that fixes the total number of samples to $N$ and a Lagrange multiplier $\lambda$,
\begin{align}
    \mathcal{L} = \sum_{i=1}^{n_c} N_i^{1-\alpha} \; \text{Var}_\alpha(\hat{I}_i) - \lambda \left( N - \sum_{i=1}^{n_c} N_i \right) \; .
\end{align}
We then optimize by setting the derivative to zero,
\begin{align}
\begin{split}
    0 &= \frac{\partial \mathcal{L}}{\partial N_i}
    = (1 - \alpha) N_i^{-\alpha} \; \text{Var}_\alpha(\hat{I}_i) + \lambda \\
    \implies \quad N_i &= \left( \frac{\alpha-1}{\lambda} \right)^{1/\alpha} \text{Var}_\alpha(\hat{I}_i)^{1/\alpha} \\
    \implies \quad N_i &= N \frac{\text{Var}_\alpha(\hat{I}_i)^{1/\alpha}}{\sum_{j=1}^{n_c} \text{Var}_\alpha(\hat{I}_j)^{1/\alpha}} \; .
\end{split}
\end{align}
Inserting this result back into the loss function from Eq.~\eqref{eq:alpha_var_bound} and normalizing to remove the scaling with the total $N$ yields the final result for the stratified $\alpha$-variance loss,
\begin{align}
\begin{split}
    L &= N^{\alpha-1} \sum_{i=1}^{n_c} N_i^{1-\alpha} \; \text{Var}_\alpha(\hat{I}_i) \\
    &= N^{\alpha-1} \sum_{i=1}^{n_c} \text{Var}_\alpha(\hat{I}_i)
    \left( N \frac{\text{Var}_\alpha(\hat{I}_i)^{1/\alpha}}{\sum_{j=1}^{n_c} \text{Var}_\alpha(\hat{I}_j)^{1/\alpha}} \right)^{1-\alpha} \\
    &= \sum_{i=1}^{n_c} \text{Var}_\alpha(\hat{I}_i)^{1/\alpha}
    \left( \sum_{j=1}^{n_c} \text{Var}_\alpha(\hat{I}_j)^{1/\alpha} \right)^{\alpha-1} \\
    &= \left( \sum_{j=1}^{n_c} \text{Var}_\alpha(\hat{I}_j)^{1/\alpha} \right)^\alpha \; .
\end{split}
\end{align}

\begin{figure}[tb]
    \centering
    \includegraphics[width=0.49\textwidth]{figs_old/alphavar_wp2j_stddev.pdf}
    \includegraphics[width=0.49\textwidth]{figs_old/alphavar_wp2j_uweff.pdf}
    \caption{Final standard deviation, final unweighting efficiency and loss functions for networks trained with an stratified $\alpha$-variance loss and different values of $\alpha$ for W+2j.}
    \label{fig:alphavar_wp2j}
\end{figure}

\section*{Acknowledgements}
OM, FM and RW acknowledge support by FRS-FNRS (Belgian National Scientific Research Fund) IISN projects  4.4503.16. TP would like to thank the Baden-W\"urttemberg-Stiftung for financing through the program \textit{Internationale Spitzenforschung}, project
\textsl{Uncertainties --- Teaching AI its Limits}
(BWST\_IF2020-010). TP is supported by the Deutsche
Forschungsgemeinschaft (DFG, German Research Foundation) under grant
396021762 -- TRR~257 \textsl{Particle Physics Phenomenology after the
Higgs Discovery}. TH is funded by the Carl-Zeiss-Stiftung through the
project \textsl{Model-Based AI: Physical Models and Deep Learning for
Imaging and Cancer Treatment}.
The authors acknowledge support by the state of Baden-Württemberg through bwHPC and the German Research Foundation (DFG) through grant no INST 39/963-1 FUGG (bwForCluster NEMO). Computational resources have been provided by the supercomputing facilities of the Université catholique de Louvain (CISM/UCL) and the Consortium des Équipements de Calcul Intensif en Fédération Wallonie Bruxelles (CÉCI) funded by the Fond de la Recherche Scientifique de Belgique (F.R.S.-FNRS) under convention 2.5020.11 and by the Walloon Region.

\clearpage

\bibliography{tilman,refs}